\documentclass[reprint,superscriptaddress,showpacs,amssymb,amsmath,aps,prd,longbibliography]{revtex4-1}
\usepackage{graphicx}
\usepackage{xcolor}
\usepackage{amsmath}
\usepackage{amsthm}
\usepackage{bm}
\usepackage{bbm}
\usepackage{mathdots}
\usepackage{lipsum}
\usepackage{verbatim}
\usepackage[caption=false]{subfig}
\newcommand{\T}{\text{\footnotesize\ensuremath{T}}}

\usepackage[mathscr]{euscript}

\begin{document}
	
	\title{Space-time Quantum Actions}
	
	\author{N. L.\  Diaz}
	\affiliation{Departamento de F\'isica-IFLP/CONICET,
		Universidad Nacional de La Plata, C.C. 67, La Plata (1900), Argentina}
	
	\author{J. M. Matera}
	\affiliation{Departamento de F\'isica-IFLP/CONICET,
		Universidad Nacional de La Plata, C.C. 67, La Plata (1900), Argentina}
	
	\author{R. Rossignoli}
	\affiliation{Departamento de F\'isica-IFLP/CONICET,
		Universidad Nacional de La Plata, C.C. 67, La Plata (1900), Argentina}
	\affiliation{Comisi\'on de Investigaciones Cient\'{\i}ficas (CIC), La Plata (1900), Argentina}

	\begin{abstract} 
		We propose a formulation of quantum mechanics in an extended Fock space in which a tensor product structure is applied to time. Subspaces of histories consistent with the dynamics of a particular theory are defined by a direct quantum generalization of the corresponding classical action.  
		The diagonalization of such quantum actions enables us to recover the predictions of conventional quantum mechanics and reveals an extended unitary equivalence between all physical theories. 
		Quantum correlations and coherent effects across time and between distinct theories acquire a rigorous meaning, which is encoded in the rich temporal structure of physical states. Connections with modern relativistic schemes and the path integral formulation also emerge.
	\end{abstract}
	\maketitle
	
	\section{Introduction}
	Quantum mechanics (QM) is a mathematical framework for the development of physical theories \cite{nie.01}. This framework assigns an operator acting on a Hilbert space for each observable of a given system, e.g. the position of a particle. In particular, the Hamiltonian operator corresponds to the energy of the system and determines its quantum evolution, defining thus the particular theory. On the other hand, the spectral properties of a general Hamiltonian preclude the introduction of a time operator, a result known as Pauli's theorem \cite{Pau.90,lm.20,col.19}: In the canonical formulation of QM time is treated ``clasically'', i.e. it is not part of the framework as an observable \cite{lm.20}.

	This manifest asymmetry between space and time is in clear contrast with the covariance of classical (relativistic) physics, a problem partially overcome in canonical formulations of relativistic quantum field theories: Classical theories are quantized on a time-slice \cite{dy.94} and 
	space becomes an index indicating the site of an ``oscillator''. In this way, transformations mixing space and time (e.g. Lorentz transformations) can be introduced. However, since the latter is an external parameter, not the index of a site, 
	at the Hilbert space level an asymmetry is still present \cite{di.19,dia.19}: A tensor product structure is applied to space but not to time, as observed in \cite{ish.93,Ish.98,fit.15,ho.17,zh.18,cot.18, bru.19}.
	This is a manifestation of fundamental open problems concerning the proper treatment of general covariance on Hilbert space \cite{ish.93,we.20,chat.20,g.09,ku.11,boj.11},  which are an important motivation for the recent interest on the introduction of time in a purely quantum framework \cite{QT.15,fit.15,b.16,ho.17,zh.18,b.18,nik.18,m.18,cot.18,sm.19, bru.19,di.19,dia.19,lm.20,we.20,chat.20,he.20,pla.20}. However, the asymmetry is present in any composite system \cite{ho.17,fit.15}. In particular, this prevents the representation of trajectories in a Hilbert space (see Sec.\ \ref{II.A}) and the  use of conventional tools for describing  quantum correlations in time \cite{megidish2013,pabon2019parallel}.
	
    In this work, the conventional framework of QM is generalized to remove the 
    above-stated
    asymmetry. This is accomplished by formulating quantum mechanics in an extended Fock space in which a tensor product structure is applied to time (previous attempts in this direction include \cite{Ish.98}, see discussion in Sec.\ \ref{II.A}). 
    The formalism is presented in section \ref{II}, together with  the concept of space-time quantum actions and the  definition of physical states. The case of quadratic theories is analyzed in detail in section \ref{III},  where   connections with other formalisms through second quantization and relativistic considerations are also examined.
	Different proposals for obtaining physical predictions in the general case within the present extended framework, 
  including states at a given time through quantum foliation and path integrals, 
  are discussed in section \ref{IV}. A final discussion is provided in section \ref{V}.

	\section{Formalism\label{II}}
	\subsection{A Hilbert Space for Quantum Trajectories}\label{II.A}

\begin{figure}[ht]
		\hspace{-0.13cm}
		\includegraphics[width=0.4825\textwidth]{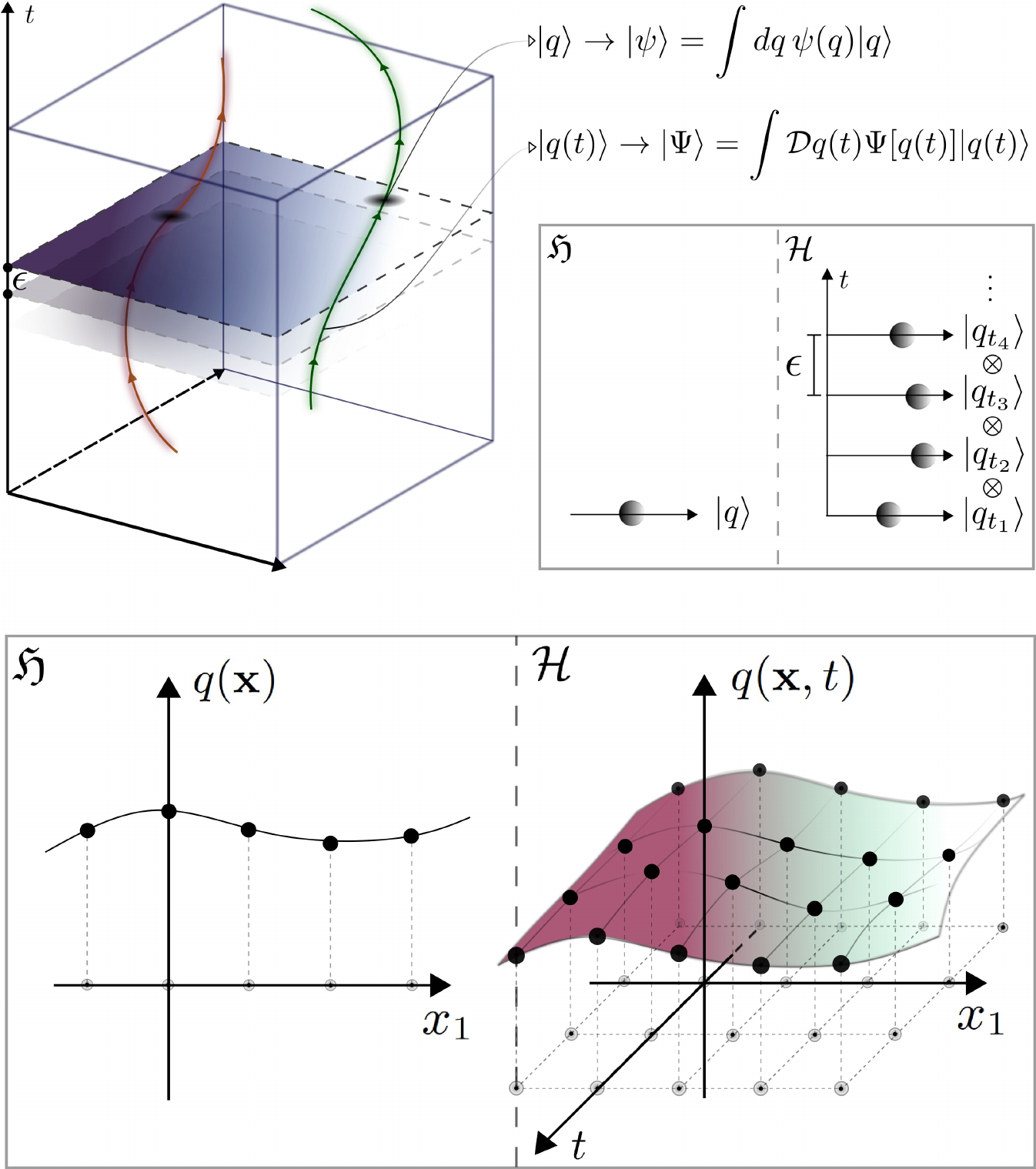}
		\caption{Representation of two classical (distinguishable)  particles moving in flat space-time whose trajectories can be parameterized as $(t,q_a(t),q_b(t))$ (top left). Conventional QM describes this situation by employing a basis of product states $|\bm q\rangle=|q_a\rangle\otimes|q_b\rangle$ which represent the positions at a given time in the Hilbert space $\mathfrak{H}$.  Instead, in $\mathcal{H}$ the whole paths are represented by
			$|\bm q(t)\rangle=|q_a(t)\rangle \otimes |q_b(t)\rangle\propto\bigotimes_j|q_{at_j}\rangle\otimes
			|q_{bt_j}\rangle$ (Eq.\ \eqref{eq:qt}), 
			where $|q_i(t)\rangle\propto\bigotimes_j | q_{it_j}\rangle$
			(top-right) which establishes a completely symmetric application of the tensor product to spatial and temporal degrees of freedom.  Moreover, classical time evolution $\bm q(t)\to \bm q(t+\Delta t)$ can be seen from a passive point of view as a displacement $t\to t-\Delta t$ of the whole manifold. In our formulation, quantum time evolution emerges from $e^{i \mathcal{P}_t(-\Delta t)}|\bm q(t)\rangle=|\bm q(t+\Delta t)\rangle$. The symmetry between space and time is further depicted on the bottom panel with a different example: The tensor product in space of a conventional quantum field theory is here extended to space-time.}
		\label{fig:1}
	\end{figure}

	We introduce in this section a Hilbert space $\mathcal{H}$ suited for representing trajectories (see Fig.\ \ref{fig:1}) of a set of bosons defined by operators $a_i$, $a^\dag_j$, $[a_i,a^\dag_j]=\delta_{ij}$, $[a_i,a_j]=0$, for $i,j$ arbitrary quantum numbers (e.g. $i$ may represent a  discretized position  $\textbf{x}$), which  generate a ``conventional'' Fock space $\mathfrak{H}$ of states $\prod_i(a^\dag_i)^{n_i }|0\rangle$ (with $a_i|0\rangle=0$). For this purpose we define 
	creation/annihilation operators $A_i(t)$, $A^\dag_j(t)$ on ``each'' time-slice, satisfying $[A_i(t),A_j(t')]=0$ and  
	\begin{equation}\label{eq:alg}
	[A_i(t),A_j^\dag(t')]=\delta(t-t')\delta_{ij}\,,
	\end{equation} 
	with $A_i(t)|\Omega\rangle=0$ 
	$\forall t\in[-\T/2,\T/2]$, which generate an extended Fock space 
	${\cal H}$. Here  $|\Omega\rangle=\bigotimes_j|0\rangle_{t_j}$,  where the tensor product is to  be interpreted as the continuum limit of equally spaced discrete time ``sites''   with spacing $\epsilon$,  
	 such that $t_j=\epsilon j$, $j\in \mathbb{Z}$ and  $A_i(t_j)=A_{it_j}/\sqrt{\epsilon}$, 
	 with $A_{it_j}|0\rangle_{t_j}=0$ and  $[A_{it_j},A^\dag_{i't_{j'}}]=\delta_{jj'}\delta_{ii'}$.  The algebra of Eq.\ (\ref{eq:alg}) is recovered from $\delta(t_j-t_{j'})\equiv\delta_{jj'}/\epsilon$. 
	 
	 The extended Hilbert space 
	 ${\cal H}$ 
	 of 
	 states $\prod_{i,j}(A^{\dag}_{i t_j})^{n_{ij}}
	 |\Omega\rangle$ can then be written as $\mathcal{H}=\bigotimes_j\mathfrak{H}_{t_j}$ with $\mathfrak{H}_{t_j}$ the Fock space generated by the operators $A^\dag_{it_j}$ (fixed $j$). Note also that we can write $\mathfrak{H}=\bigotimes_i \mathfrak{H}_i$ and then 
	\begin{equation*}
	\mathcal{H}=\bigotimes_{i,j}\mathfrak{H}_{ij}\,,
	\end{equation*}
	with $\mathfrak{H}_{ij}\equiv \mathfrak{H}_{it_{j}}$, which is the aimed Hilbert space symmetry between ``space'' (index $i$) and time (see Fig.\ \ref{fig:1}).
	
	This construction allows us to specify, up to quantum uncertainty, a classical \emph{trajectory} in phase space as a coherent history state, i.e. a product state of the form 
	\begin{equation}\label{eq:coherentst}
	|\bm{\alpha}(t)\rangle:=\exp\left[\int dt
	\,\bm{\alpha}(t)\cdot\bm{A}^\dag(t)\right]|\Omega\rangle\,,
	\end{equation}
	where $\bm{\alpha}(t)\cdot  \bm{A}^\dag(t)=\sum_{i} \alpha_i(t) A_i^\dag(t)$ (or an integral for continuum labels). 

	Here $\exp\left[
	\int\,dt\, \mathcal{O}(t)\right]=\bigotimes_j \exp\left[\epsilon\,\mathcal{O}(t_j)\right]$, 
	where $\mathcal{O}(t)\equiv \mathcal{O}(\bm{A}(t),\bm{A}^\dag(t),t)$, such that 
	\begin{equation}
	\bm{A}(t)|\bm{\alpha}(t)\rangle=\bm{\alpha}(t)|\bm{\alpha}(t)\rangle\,.
	\end{equation}
	Note that 
	$|\bm{\alpha}(t)\rangle=e^{\alpha A^\dag}|\Omega\rangle$, where  $A^\dag=\!\int\! dt \bm\alpha(t)\cdot\bm{A}^\dag(t)/\alpha$ with   $\alpha=[\int dt |\bm\alpha(t)|^2]^{1/2}$, is a ``collective'' trajectory boson creation operator. The (over)complete set of these trajectories span $\mathcal{H}$:
	
	\begin{equation}
	\int \mathcal{D}^2\bm{\alpha}(t)\,e^{-\int dt\,|\bm{\alpha}(t)|^2}|\bm{\alpha}(t)\rangle \langle \bm{\alpha}(t)|=\mathbbm{1}
	\end{equation}
	where  $\mathcal{D}^2\bm{\alpha}(t):=\prod_{i,j}\frac{d^2\alpha_{i}(t_j)}{\pi}\epsilon$.
	
	Alternative basis  are provided for example by operators $\bm{Q}(t)=\tfrac{\bm{A}(t)+\bm{A}^\dag(t)}{\sqrt{2}},\, \bm{P}(t)=\tfrac{\bm{A}(t)-\bm{A}^\dag(t)}{i\sqrt{2}}$, such that
	\begin{equation}
	    [Q_i(t),P_j(t')]=i\delta(t-t')\delta_{ij}
	\end{equation}
	 (we set $\hbar=1$). 
	Then we can define the corresponding eigenstates $|\bm{q}(t)\rangle$, 
	$|\bm{p}(t)\rangle$, satisfying   
		\begin{equation}\label{eq:qpt}
	\bm{Q}(t)|\bm{q}(t)\rangle=
	\bm{q}(t)|\bm{q}(t)\rangle\,,\;\; \bm{P}(t)|\bm{p}(t)
	\rangle=\bm{p}(t)|\bm{p}(t)\rangle\,.
	\end{equation}
	Explicitly, we can write  \footnote{The canonical position eigenstate $|q\rangle$ can be expressed as $|q\rangle=\sum_{n=0}^\infty \langle n|q\rangle \frac{(a^\dag)^n}{\sqrt{n!}}|0\rangle=\frac{e^{-q^2/2}}{\sqrt[4]{\pi}} e^{-\frac{a^{\dag\,2}}{2}+\sqrt{2}q a^\dag}|0\rangle$ \cite{soto.13}. The multi-dimensional case 
		$|\bm{q}\rangle=\bigotimes_i \frac{e^{-q_i^2/2}}{\sqrt[4]{\pi}} e^{-\frac{a_i^{\dag\,2}}{2}+\sqrt{2}q_i a_i^\dag}|0\rangle\propto  e^{-\tfrac{1}{2}[\sum_ia_i^{\dag} (a_i-2\sqrt{2}q_i)}|0\rangle$ can be immediately generalized to continuum fields \cite{qft.14}.},   \cite{soto.13,qft.14} 
	\begin{eqnarray}\label{eq:qt}
	|\bm{q}(t)\rangle&=&
	\exp[-\tfrac{1}{2}\int dt\,
	\bm{A}^\dag(t)\cdot(\bm{A}^\dag( t)-2\sqrt{2}\bm{q}(t))]|\Omega\rangle \;\;\;
\end{eqnarray}
 such that $|\bm{q}(t)\rangle=\bigotimes_j \gamma_j|\bm{q}_{t_j}\rangle_{t_j}$ with $\bm{q}_{t_j}=\sqrt{\epsilon}\bm{q}(t_j)$,  $\gamma_j=\sqrt[4]{\pi}\,e^{|\bm{q}_{t_j}|^2/2}$ and $_{t_j}\langle \bm{q}_{t_j}|\bm{q}'_{t_{j}}\rangle_{t_j}=
 \delta(\bm{q}-\bm{q}')$.  The completeness relation reads 
  $\int{\cal D}\bm{q}(t)\,e^{-\int dt\,|\bm{q}(t)|^2} |\bm{q}(t)\rangle\langle \bm{q}(t)|=\mathbbm{1}$  (${\cal D}\bm{q}(t)=\prod_{i,j}
	d q_i(t_j)\sqrt{\pi \epsilon}$). Similar formulas hold for $|\bm{p}(t)\rangle$. These space-time bases enable a novel approach for path integral representations,   as will be discussed in Sec.\ \ref{S:PI}.  
	
	While $\mathcal{H}$ is isomorphic to a tensor product of copies in time of $\mathfrak{H}$, we have not specified any particular time evolution yet. We have only introduced a suitable  ``geometrical'' scenario (which may be indicated as \emph{space-time}) in which any laws of physics may be defined. 
	In fact, a ket in $\mathcal{H}$ does not ``evolve'' but it can contain by itself all the time information (or history) of a given system. Some condition must establish which ones of these histories is compatible with a particular theory, an intuition which leads  us to the definition of \emph{physical subspaces} $\mathcal{H}_P$. 
	It turns out that if we propose that the trivial theory (null Hamiltonian) is defined by those coherent states invariant under time translations, a natural definition for all theories follows. This result, which is presented in Sec.\ \ref{sec:pstates}, relies on the extended unitary equivalence between theories that we introduce in Sec.\ \ref{sec:timetrans}. 

	We also note that a similar discrete tensor product in time  Hilbert space is employed in the context of the `consistent-histories'  approach to quantum mechanics 
	introduced by Isham \cite{ish.93}, with the aim of providing a novel way of representing the corresponding decoherence functional. The latter  is the central quantity in the scheme developed in \cite{grif.84,gellm.19}, concerning  the joint probability of finding  a sequence
of properties at a series of times. In Isham's approach, a copy of the original Hilbert space is involved for each of these times. In its continuous-time formulation \cite{Ish.98}, the  basic operators also satisfy Eq.\ \eqref{eq:alg}.
Nevertheless, in the present formalism, this enlarged Hilbert space, rather than a tool for representing histories, is considered as fundamental.
In particular, time evolution is \emph{derived} from properties of the corresponding time translation operator and encoded in physical states while the ``number of time sites'' is arbitrary. Quantities such as the decoherence functional can be obtained a posteriori.

	\subsection{Time Translations and Space-time Quantum Actions}\label{sec:timetrans}
	Consider 
	the generator of time translations $\mathcal{P}_t$ in the present scenario,  	defined as
	\begin{subequations}
	\begin{eqnarray}
	\mathcal{P}_t&:=&\int d
	\omega\, \omega\bm{A}^\dag(\omega)\cdot\bm A(\omega)\\&=&\int dt\, \bm{A}^\dag(t)\cdot i\bm{\dot{A}}(t)\label{ptadot}\\
	&=&\tfrac{1}{2}\int dt\, [\bm{P}(t)\cdot \bm{\dot{Q}}(t)-\bm{Q}(t)\cdot \bm{\dot{P}}(t)]
	\label{PQ}	\end{eqnarray}
	\label{Ptg}
	\end{subequations}
	where  $\bm{A}(\omega)$ is the Fourier transform (FT) of $\bm{A}(t)$, such that   $\bm A(t)=
	\int  \tfrac{d\omega}{\sqrt{2\pi}}\bm A(\omega)e^{-i\omega t}$ (continuous notation, see Appendix \ref{ApA})   
	and 
	$i\bm{\dot{A}}(t)=\int  \tfrac{d\omega}{\sqrt{2\pi}}\bm A(\omega)\,\omega e^{-i\omega t}$ coincides with the ``site'' derivative (Eq.\ \eqref{A3}). We assume  periodic conditions $\bm{A}(-\T/2)=\bm{A}(\T/2)$. The  operator ${\cal P}_t$ satisfies 
	\begin{equation}
	e^{i\mathcal{P}_t\Delta t}\bm{A}(t)e^{-i\mathcal{P}_t\Delta t}=\bm{A}(t+\Delta t)\,,\label{eq:pt}\end{equation}
	which for $\Delta t\rightarrow 0$ leads to 
	\begin{equation}\label{eq:ptconm} [{\cal P}_t,\bm{A}(t)]=-i\bm{\dot{A}}(t)
\,,
	\end{equation}
	in agreement with Eq.\ (\ref{ptadot}). 
	
	Remarkably, the integrand in \eqref{PQ}
	has the form of the \emph{Legendre transformation} which connects the Hamiltonian with the Lagrangian in classical mechanics.  This suggests the introduction of a new object that for the trivial theory reduces to $\mathcal{P}_t$:
	\begin{equation}\label{eq:action}
	\mathcal{J}:=\int dt\, [ \bm{A}^\dag(t)\cdot i\bm{\dot{A}}(t)-\mathscr{H}(\bm{A}(t),\bm{A}^\dag(t),t)\,,
	\end{equation}
	which will be indicated as  \emph{space-time quantum action operator} (not to be confused with Schwinger's action \footnote{In Schwinger's formulation, 
		a complete set of commuting operators is available on space-like surfaces \cite{sch.51}. This implies non-vanishing commutators for causally connected field operators. Instead, any unequal-time commutator between $\bm{A}(t)$ and $\bm{A}^\dag(t)$ vanishes. Conventional algebras are recovered ``a posteriori'' in the physical subspaces. Moreover, the integration in $\mathcal{J}$ involves all values of time without any reference to particular states.}, \cite{sch.51})
	for its formal coincidence with the classical one. Here 
	$\int dt\, \mathscr{H}(\bm{A}(t),\bm{A}^\dag(t),t)\equiv \sum_t H(\bm{A}_t,\bm{A}^\dag_t,t)$ for $H(\bm{a},\bm{a}^\dag,t)$ a conventional (quantum)
	Hamiltonian (and $dt=\epsilon$), in accordance with the convention of $\mathcal{J}$ having units of $\mathcal{P}_t$. 
	A remarkable result is that $\mathcal{J}$ and
	$\mathcal{P}_t$ are {\it unitarily related}  (see proof in Appendix \ref{ApB}): 
	\begin{subequations}
	\label{Jgral}
	\begin{eqnarray}\label{eq:jdiag}
		\mathcal{J}=\mathcal{V}^\dag \mathcal{P}_t \mathcal{V}&=&\int d
	\omega\, \omega\bm{\tilde{A}}^\dag(\omega)\cdot\bm{\tilde{A}}(\omega)\\&=&\int dt\, \bm{\tilde{A}}^\dag(t)\cdot i\bm{\dot{\tilde{A}}}(t)\label{eq:Jat}\\&=&\tfrac{1}{2}\int dt\, [\bm{\tilde{P}}(t)\cdot \bm{\dot{\tilde{Q}}}(t)-\bm{\tilde{Q}}(t)\cdot \bm{\dot{\tilde{P}}}(t)]\label{eq:Jpq}
	\,,\;\;\;\;\;\;
		\end{eqnarray}
		\end{subequations}
	where 
	\begin{equation}\label{eq:unitary}
	\mathcal{V}^\dag:=\hat{T}' \exp\Bigl[-i\int dt\,\int_{t_0}^{t}dt'\, \mathscr{H}(\bm{A}(t),\bm{A}^\dag(t),t') \Bigr]
	\end{equation}
 is a {\it tensor product in time of conventional time evolution operators} $U(t,t_0)=\hat{T}'\exp[-i\int_{t_0}^{t} dt'\,H(\bm{a},\bm{a}^\dag,t')]$ 
 ($\hat{T}'$ denotes 
	time ordering  
	applied to $t'$)  and 
 \begin{equation}
 \bm{\tilde{A}}(\omega)=\mathcal{V}^\dag \bm{A}(\omega) \mathcal{V}\,,\;\;\;\;    \bm{\tilde{A}}(t)=\mathcal{V}^\dag \bm{A}(t) \mathcal{V}\,,\label{eq:Atild}
 \end{equation} with $\tilde{\bm{A}}(t)$ the FT of $\bm{\tilde{A}}(\omega)$ 
	(similarly $\bm{\tilde{Q}}(t)={\cal V}^
\dag \bm{{Q}}(t){\cal V}$, $\bm{\tilde{P}}(t)={\cal V}^
\dag \bm{{P}}(t){\cal V}$).   Here $t_0$ is a reference time such that  $\bm{\tilde{A}}(t_0)=\bm{A}(t_0)$. 
	In particular, for $\mathscr{H}$ time independent,
	\begin{equation}\label{eq:VH0}
	\mathcal{V}^\dag=\exp[-i \int dt\, (t-t_0) \mathscr{H}(\bm{A}(t),\bm{A}^\dag(t))]\,.\end{equation}
	
 Since in this context $\mathcal{J}$ is the operator that defines a particular time evolution (Sec.\ \ref{sec:pstates}), the result (\ref{eq:jdiag}) is unitarily relating all theories to the trivial one. This also means that in $\mathcal{H}$ all physical theories appear unitarily related between themselves.	Such general result is a consequence of the remarkable property of the space-time quantum actions of having the same spectra regardless of the Hamiltonian. This should be compared with the obvious fact that different Hamiltonians have different spectra, which also means that such unitary relation between theories could have never been revealed in a Hamiltonian formulation.

		The proof of  (\ref{eq:jdiag}) is based on the basic properties of $\mathcal{P}_t$ as the generator of time translations, and assumes periodic conditions for finite $\T$ (something which in principle can always be ``enforced'' or  implemented by a ``well behaved'' $\mathscr{H}$ in the limit $\T\to \infty$). 
		Notice that Eqs.\ \eqref{eq:pt} and  \eqref{Jgral} entail 
	\begin{equation}e^{i\mathcal{J}\Delta t}\bm{\tilde{A}}(t)e^{-i\mathcal{J}\Delta t}=\bm{\tilde{A}}(t+\Delta t)\label{eq:Jpt}\end{equation}
	such that ${\cal J}$ is the generator of time translations in the ``normal'' basis for a non-null Hamiltonian. Therefore, the operators $\bm{\tilde{A}}(t)$ satisfy 
	\begin{equation}
	[\mathcal{J},\bm{\tilde{A}}(t)]=-i\bm{\dot{\tilde{A}}}(t)\,,\label{eq:conm}
	\end{equation}
	in accordance with \eqref{eq:Jat}. In fact, they are the unique annihilation operators fulfilling  
	\eqref{eq:conm} and $\tilde{\bm{A}}(t_0)=\bm{A}(t_0)$. The uniqueness is an immediate consequence of \eqref{eq:Jpt} which implies
	\begin{equation}\label{eq:ci}
	  \bm{\tilde{A}}(t)=e^{i\mathcal{J}\Delta t}\bm{A}(t_0)e^{-i\mathcal{J}\Delta t}\,
	\end{equation}
	when $\Delta t=t-t_0$. The relation (\ref{eq:ci}) is a remarkable result on its own which provides an expansion in powers of $\Delta t$ of the ``evolved'' operator $\mathcal{V}^\dag \bm{A}(t)\mathcal{V}$ (see also Appendix \ref{ApB} and the discussion below). In the context of the consistent histories approach, and for  the particular  case of a time-independent  harmonic oscillator, an  analogous action complying with Eq.\ \eqref{eq:ci} was introduced in \cite{Sav.99}.    
	
	Before proceeding to the definition of physical subspaces, we would like to stress that as a consequence of (\ref{Jgral})--(\ref{eq:Atild}) the information of conventional time evolution is already encoded in the operators $\bm{\tilde{A}}(t)$:
	From Eq.\ (\ref{eq:unitary}) it is clear that the operator $\tilde{\bm{A}}(t)$ 
	corresponds to the operator $\bm{a}(t)=U(t,t_0)  \bm{a} U^\dag(t,t_0)$, which acts on $\mathfrak{H}_t$. Since an underlying tensor product is involved, this statement is rigorous for discrete time, in which case we can also speak properly of ``instants'' and ``sites''. In Sec.\ \ref{sec:prop} these ideas and the discrete regularization will be employed to derive (and interpret) different ways to obtain \emph{physical predictions from the inner product of} $\mathcal{H}$. On the other hand, the expressions involved can also be obtained straightforwardly in the $\omega$ basis 
	by employing the normal operators
	 $\bm{\tilde{A}}(\omega)$ of \eqref{eq:jdiag}, which satisfy 
	\begin{equation}
	    [\mathcal{J},\bm{\tilde{A}}^\dag(\omega)]=\omega\bm{\tilde{A}}^\dag(\omega)\,.\label{eq:Aw}
	\end{equation}
	In this basis the limit $\epsilon\to 0^+$ is well defined and a map with conventional states in $\mathfrak{H}$ can be easily introduced. 
	
	We also remark that 
	for a general 
	periodic (or well behaved in the limit $\T \to \infty$) operator 
	$
	{\cal U}=\exp[\int dt\,{\cal M}(\bm{A}(t),\bm{A}^\dag(t),t)]\,,
	$
	Eq.\ (\ref{eq:pt}) yields (see Appendix \ref{ApB} for the details)	\begin{equation}[{\cal P}_t,{\cal U}]=i\frac{\partial{\cal U}}{\partial t}\,
	\end{equation}
	with $\frac{\partial{\cal U}}{\partial t}$ defined in (\ref{eq:derivative}) through Eq. (\ref{eq:ptconm}). 
	For $\mathcal{M}(\bm{A}(t),\bm{A}^\dag(t))$ time independent, Eq.\ (\ref{eq:derivative}) implies $[\mathcal{P}_t,\mathcal{U}]=0$. If $i\mathcal{M}(\bm{A}(t),\bm{A}^\dag(t))$ is also hermitian, this implies $\mathcal{U}^\dag \mathcal{P}_t\mathcal{U}=\mathcal{P}_t$, i.e. $\mathcal{P}_t$ is invariant under time independent \emph{canonical transformations} $\bm{A}(t)\to \mathcal{U}^\dag \bm{A}(t)\mathcal{U}$.
 This means that without imposing any initial conditions, the diagonal form (\ref{eq:jdiag}) is not unique and implies \begin{equation}\label{eq:uh}
	[\mathcal{U},\int dt\,\mathscr{H}(\bm{A}(t),\bm{A}^\dag(t),t)]=0 \Rightarrow [\mathcal{U},\mathcal{J}]=0\,. 
	\end{equation} 
	In particular, \emph{a time-independent symmetry of $H$, 
	$[M(\bm{a},{\bm{a}}^\dag),H(t)]=0$, is a symmetry of $\mathcal{J}$}: $[{\cal U},{\cal J}]=0$, for $\epsilon\mathcal {M}(\bm{A}(t),{\bm{A}}^\dag(t))=M(\bm{A}_t,{\bm{A}}^\dag_t)$. On the other hand, for $H$ time-independent it follows from Eq. (\ref{eq:pt}) that  $e^{i\mathcal{P}_t\Delta t}$ satisfies Eq. (\ref{eq:uh}), i.e. $\mathcal{J}$ is \emph{invariant under time translations} and hence $[\mathcal{P}_t,\mathcal{J}]=0$ (see also Eq.\ \eqref{eq:partialder}). 
	In the Appendix \ref{ApB} we discuss further symmetries of $\mathcal{P}_t$ and ${\cal J}$ which are not diagonal in time, together with the possibility to generalize (\ref{eq:jdiag}) to ``exotic'' theories involving multiple-times.

	Finally, it is appropriate to mention
	that different definitions of time localization are now possible: As it happens for spatial localization in quantum field theories (QFT) with important implications on spatial uncertainty relations \cite{cel.16, bi.12}, \emph{time localization is now an emergent aspect} of the ``lattice''.
	Different definitions of this notion would also imply different \emph{energy-time uncertainty relations} according to the operators involved. An example is provided by	the single particle (sp) \emph{time operator} $\mathcal{T}:=\int dt\,t \bm A^\dag(t)\cdot\bm A(t)$ which reduces on sp states to the Page and Wootters (PaW) operator \cite{PaW.83} (see Sec.\ \ref{IV} D)  employed in other recent formalisms with quantum time  \cite{lm.20,we.20,di.19,dia.19,b.18,nik.18, m.18,sm.19,QT.15}.
	In this case, it can be shown that (see Eq.\ (\ref{eq:derivative}); here $\T\to \infty$)
	\begin{equation}
	[\mathcal{P}_t,\mathcal{T}]=i\mathcal{N}\,,\label{con}
	\end{equation} 
	where $\mathcal{N}:=\int dt\,  \bm A^\dag(t)\cdot\bm A(t)=\int d\omega\,  \bm A^\dag(\omega)\cdot\bm A(\omega)$ is the number operator (e.g.\ ${\cal N}(A_i^\dag(t))^{n_i}|\Omega\rangle=n_i(A_i^\dag(t))^{n_i}|\Omega\rangle$).  Then $\Delta \mathcal{T} \Delta \mathcal{P}_t\geq \tfrac{1}{2}|\langle \mathcal{N} \rangle|$ through 
	the Cauchy–Schwarz inequality in $\mathcal{H}$. Despite the importance of the energy-time pair in QM \cite{col.19}, this treatment is usually prevented by the impossibility of introducing a time operator in $\mathfrak{H}$ \cite{Pau.90, ah.61, col.19}. 
	
	\subsection{Physical States}\label{sec:pstates} We are now in a position to formalize the postulates that define a particular physical theory: 
	Consider the normal operators $\bm{\tilde{A}}(\omega)$ defined by the representation (\ref{eq:jdiag}) of the quantum action, fulfilling Eq.\ \eqref{eq:Aw} and $\bm{\tilde{A}}(t_0)=\bm{A}(t_0)$, and their vacuum $|\tilde{\Omega}\rangle=\mathcal{V}^\dag|\Omega\rangle$.  The corresponding ${\cal H}_P$ is introduced
	as the linear space spanned by  states $\prod_i(\tilde{A}^\dag_i(\omega=0))^{n_i}|\tilde{\Omega}\rangle$, i.e. the Fock space generated by the creation operators satisfying
	\begin{equation}\label{eq:physsubs}
	[\mathcal{J},\bm{\tilde{A}}^\dag(0)]=0\,,
	\end{equation}
	which may be interpreted as a static (or timeless) Heisenberg equation for $\bm{\tilde{A}}^\dag(0)$. This definition is in accordance with the proposal in \cite{dia.19} which originated from relativistic considerations. In particular, since just $\omega=0$ bosons are involved, 
	$\mathcal{J}|\Psi\rangle=0$ $\forall$ $|\Psi\rangle\in{\cal H}_P$,  a constraint which defines related quantum formalisms \cite{pam.50,PaW.83} motivated by the Wheeler-DeWitt equation \cite{wdw.67} (see also Sec.\ \ref{S4D}).
	Eqs.\ \eqref{eq:jdiag},  \eqref{eq:physsubs} also imply  $\langle \Psi|\frac{\delta \mathcal{J}}{\delta \bm{\tilde{A}}(\omega)}|\Psi\rangle=0$, 
	meaning that the average of the quantum action ${\cal J}$ is \emph{stationary} in $\mathcal{H}_P$ as a functional of $\bm{\tilde{A}}(\omega)$ \footnote{Here $\frac{\delta\mathcal{J}}{\delta \bm{\tilde{A}}(\omega)}
		=\omega\bm{\tilde{A}^\dag}(\omega) =[\mathcal{J},\bm{\tilde{A}}(\omega)]$, an equation which defines normal modes.}.
	
	In order to show that the present formalism yields (in a physical subspace)  the same predictions of conventional QM, we  establish an isomorphism ${\cal L}:\mathfrak{H}\rightarrow {\cal H}_P$ 
	such that 
	\begin{equation}\label{eq:identif}{\cal L} \big(\prod_i[(a^\dag_i)^{n_i}]|0\rangle\big)=\prod_i[(\tilde{A}^\dag_i(0))^{n_i}]|\tilde{\Omega}\rangle\,.\end{equation} 
	We will say that $|\Psi\rangle=\mathcal{L}({|\psi\rangle})$ is the history of $|\psi\rangle\in \mathfrak{H}$ with the Hamiltonian that defines $\mathcal{J}$. 
	In particular, for a coherent state 
	$|\psi\rangle=e^{\bm{\alpha}\cdot\bm{a}^{\dag}}|0\rangle$, 
	\eqref{eq:identif} leads to 
	\begin{eqnarray}
	|\Psi\rangle&=&\exp[{\bm{\alpha}\cdot\bm{\tilde{A}}^{\dag}(\omega=0)}]|\tilde{\Omega}\rangle
	=\exp\left[\int\!
	\!\frac{dt}{\sqrt{\T}}\,\bm{\alpha}\cdot\bm{\tilde{A}}^{\dag}(t)\right]|\tilde{\Omega}\rangle
	\;\;\nonumber\\
	&=&{\cal V}^\dag\exp\left[{\int\! \frac{dt}{{\sqrt{\T}}}\bm{\alpha}\cdot\bm{A}^\dag(t)}\right]|\Omega\rangle\,,\label{cohg}
	\end{eqnarray}
	which is a product of evolved states when $\mathcal{V}^\dag$ is the operator (\ref{eq:unitary}).
	Thus,  the time invariance proposed for history coherent states of the trivial theory  (${\mathscr H}=0$, ${\cal V}^\dag=\mathbbm{1}$) unitarily defines any other. 
	An important property follows from 
	\eqref{eq:identif}: If $|\Phi\rangle$ is the history of $|\varphi\rangle$ then
	\begin{equation}\label{eq:product}
	\langle \Phi|\Psi\rangle=(\mathcal{L}(| \varphi\rangle),\mathcal{L}(|\psi\rangle))=\langle \varphi|\psi\rangle\,,
	\end{equation}
	and in particular $\langle \Psi|\Psi\rangle=\langle \psi|\psi\rangle$, a relation which holds for any $\T$, as it follows from $[\tilde{A}_i(0),\tilde{A}_j^\dag(0)]=\delta_{ij}$. Moreover, even if an infinite extent of time is considered, a natural approach emerges: The formalism treats $\omega$ as a usual continuous quantum number with an associated eigenfunction expansion. This may be regarded as an eigenbasis associated with different physical theories labeled by $\omega$: A state can be normalized if a \emph{quantum uncertainty in the physical theory} is allowed (see App.\ \ref{ApC}).

	\section{The quadratic case\label{III}}
	\subsection{Quadratic Space-time Quantum Actions}
	In the following, we explicitly develop the case of bosonic quadratic theories as an important example of (\ref{eq:action}). 
	For a general quadratic Hamiltonian
	\footnote{\label{ftqh} For $q_i=(a_i+a^\dag_i)/\sqrt{2}$, $p_i=i(a^\dag_i-a_i)\sqrt{2}$, $[q_i,p_j]=i\delta_{ij}$, $H=\frac{1}{2}\sum_{i,j}t_{ij}p_ip_j+v_{ij}q_iq_j+u_{ij}(q_ip_j+p_jq_i)$ with $t,v$ symmetric matrices,  and the matrices $t,v,u$ straightforwardly related to $\omega_0,\gamma$
		\cite{ros.05}.},\cite{ros.05}
	\begin{equation}
	H(\bm{a},\bm{a}^\dag)=\tfrac{1}{2}\begin{pmatrix}\bm a^\dag &\bm a	\end{pmatrix}\begin{pmatrix}\omega_0(t)&\gamma(t)\nonumber\\ \gamma^*(t)&\omega_0^*(t)\end{pmatrix}\begin{pmatrix}\bm a\\\bm a^\dag\end{pmatrix}=
	\tfrac{1}{2}\bm{\psi}^\dag K(t)\bm{\psi}
	\end{equation}
	where $\omega_0$ ($\gamma$) are hermitian (symmetric) matrices and  $\bm{\psi}=(^{\,\bm a}_{\bm a^\dag})$ satisfies 
	\[\Pi=[\bm{\psi},\bm{\psi}^\dag]:=\bm{\psi}\bm{\psi}^\dag-((\bm{\psi}^\dag)^t\bm{\psi}^t)^t=\begin{pmatrix}\mathbbm{1}&0\\0&-\mathbbm{1}\end{pmatrix}\,,\]
	the quantum action  \eqref{eq:action} becomes 
	\begin{equation}\label{eq:quadraticj}
	{\cal J}=\tfrac{1}{2}\int dt[ \bm{\Psi}^\dag(t)\Pi i\dot{\bm{\Psi}}(t)- \bm{\Psi}^\dag(t)K(t)\bm{\Psi}(t)]\,,
	\end{equation}
	with $\bm{\Psi}(t)=(\bm A(t), \bm A^\dag(t))^t$, $[{\bm{\Psi}}(t),{\bm{\Psi}}^\dag(t')]=\Pi\delta(t-t')$. 
	It is first verified that under any constant Bogoliubov transformation (BT) $\bm{\Psi}(t)\to W_0{\bm{\Psi}}(t)$,  where  ${W}^\dag_0\,\Pi\,W_0=\Pi$ 
	(linear time independent canonical transformation),
	the form of $\mathcal{J}$ is preserved 
	(with $K\to W^\dag_0{K} W_0$). 
	It is then seen that the diagonal 
	form (\ref{eq:jdiag}) 
	\begin{equation}
	{\cal J}=\tfrac{1}{2}\int dt\, \bm{\tilde{\Psi}}^\dag(t)\Pi i\dot{\bm{\tilde{\Psi}}}(t)=\tfrac{1}{2}\int d\omega\, \omega\, \tilde{\bm{\Psi}}^\dag(\omega)\tilde{\bm{\Psi}}(\omega)\,,
	\end{equation}
	can be
	achieved by applying in \eqref{eq:quadraticj}  a 
	\emph{diagonal in time} BT \begin{equation}
	\label{eq:BT}\bm{\Psi}(t)={ W}(t)\tilde{\bm{\Psi}}(t)\,,\end{equation} 	where $W(t)$ satisfies  
	the Heisenberg equation \footnote{\label{ft3} We assume now that  $\Pi {K}(t)$ has real eigenvalues,  which is ensured by ${K}(t)$ positive definite \cite{ros.05}, in order to warrant  periodic conditions.}
	\begin{equation}\label{eq:heisen}
	i\dot{ W}(t)=\Pi{K}(t){W}(t)\,
	\end{equation}
	with $W(t_0)=\mathbbm{1}$ in order that $\tilde{\bm{\Psi}}(t_0)=\bm{\Psi}(t_0)$ 
	(implying $W^\dag(t)\Pi W(t)=\Pi$ $\forall t$). 
	This is in agreement with  Eqs.\ (\ref{eq:unitary})--\eqref{eq:Atild} since in the present case ${\cal V}=\exp[\frac{i}{2}\int dt\bm{\Psi}^\dag(t)M(t)\bm{\Psi}(t)]$ 
	with  $e^{-i\Pi{M}(t)}={W}(t)$, and 
	\begin{equation}\label{eq:map}
	{\cal V}^\dag \bm{\Psi}(t){\cal V}=\tilde{\bm{\Psi}}(t),\;\;{\cal V}^\dag \bm{\Psi}(\omega){\cal V}=\tilde{\bm{\Psi}}(\omega)\,\end{equation}
	are BTs equivalent to \eqref{eq:BT}.  

This is the only  solution satisfying the initial condition
	 $\bm{\tilde{A}}(t_0)=\bm{A}(t_0)$, as we proved in Eq. (\ref{eq:ci}).

	\subsection{Time Structure of Physical States}
	It is important to remark that 
	the states $|\Psi\rangle\in {\cal H}_P$ constructed with Eq.\  (\ref{eq:identif}) 
	already contain all time information of the system, in a nontrivial way. 
	In fact, general physical states $|\Psi\rangle={\cal L}(|\psi\rangle)$ have a complex time structure and in particular exhibit in general {\it entanglement in time},  even for decoupled  oscillators: By considering $H=\sum_i \omega^i_0 (a_i^\dag a_i+\tfrac{1}{2})$ \footnote{Any time-independent stable quadratic Hamiltonian can be written in this diagonal normal form by an adequate choice of operators $a_i$, or equivalently, by a constant BT  $\bm{\Psi}(t)\rightarrow W_0\bm{\Psi}(t)$.} Eq.\ (\ref{eq:quadraticj}) becomes  \begin{equation}\label{eq:josc}
	{\cal J}=\sum_i\int d\omega\, (\omega-\omega^i_0)\,(A_i^\dag(\omega)A_i(\omega)+\tfrac{1}{2})\,
	\end{equation}
	such that $\tilde{A}_i(\omega)=A_i(\omega+\omega^i_0)$ in \eqref{eq:jdiag} and $\tilde{A}_i(t)=e^{i\omega_0^i t}A_i(t)$, in agreement with \eqref{eq:BT}--\eqref{eq:heisen}. 
	Then a sp state \begin{equation}\tilde{A}^\dag_i(\omega=0)|\Omega\rangle=\int \tfrac{dt}{\sqrt{\T}}e^{iw^i_0t}A_i^\dag(t)|\Omega\rangle
	=\int \tfrac{dt}{\sqrt{\T}} e^{iw^i_0t}|ti\rangle \label{eq:sppw}\end{equation} 
	is an  $W$-like  state in the time representation (unlocalized in time), where we have written  $|ti\rangle= A^\dag_i(t)|\Omega\rangle$. A general sp physical state then has the formal appearence of a PaW state \cite{b.18,lm.20} (see also Sec.\ \ref{S4D}) $|\Psi\rangle=\int \tfrac{dt}{\sqrt{\T}}\sum_i \psi_i e^{iw^i_0t}|ti\rangle$. 
	However, more general Fock states, e.g.
	\begin{equation}
	(\tilde{A}_i^\dag(0))^2|\Omega\rangle=\int \frac{dt_1}{\sqrt{\T}}\frac{dt_2}{\sqrt{\T}}e^{iw^i_0t_1}e^{iw_0^it_2} A^\dag_i(t_1)A_i^\dag(t_2)|\Omega\rangle\,,
	\end{equation}
	have even a richer structure. 
	
	On the other hand, an initial coherent state leads to coherent product  state 
(Eqs.\ \eqref{eq:coherentst} and \eqref{cohg})
	\begin{equation}
	    \mathcal{L}\left(|\bm{\alpha}\rangle\right)=|\bm{\alpha}(t)\rangle=\bigotimes_{i,j} \exp\left[\frac{\alpha_ie^{i\omega^i_0t_j}}{\sqrt{\T/\epsilon}}A^\dag_{it_j}\right]|\Omega\rangle
	\end{equation}
	i.e. $(\bm{\alpha}(t))_i=\tfrac{(\bm{\alpha})_i}{\sqrt{T}}e^{iw_0^it}$, implying 
	\begin{equation}
	    \mathcal{L}\left(\int\! \prod_i \tfrac{d^2\alpha_i}{\pi}\psi(\bm{\alpha})|\bm{\alpha}\rangle\right)=\int\! \prod_i \tfrac{d^2\alpha_i}{\pi}\psi(\bm{\alpha})|\bm{\alpha}(t)\rangle\,.
	\end{equation}
	We conclude that the physical subspace of 
	time-independent stable quadratic systems 
	corresponds to the linear space of quantum trajectories $|\bm{\alpha}(t)\rangle$, where $\bm{\alpha}(t)$ is a solution of the classical equations of motion.
	These ``almost'' classical trajectories also have a ``classical time structure'', namely \emph{separability in time}, which is an appealing property.  Remarkably, 
	$\mathcal{L}(|\psi\rangle)$ has the same formal expansion of $|\psi\rangle$ in this basis, although 
	notice that such superposition of separable (but composite) states 
	will in general be entangled.
	
	\subsection{Physical Predictions
	}\label{sec:predictionsquad}
Physical operators defined by Eq.\ (\ref{eq:map})
	satisfy, for $K$ time independent
	 ($\Delta t=t-t_0$) \begin{equation} \label{eq:quadraticev}
	e^{i \mathcal{P}_t t}\tilde{\bm{\Psi}}(0)e^{-i \mathcal{P}_t  \Delta t}=\exp(-i\Pi{K}\Delta t)\tilde{\bm{\Psi}}(0)\,,
	\end{equation}
	where $\tilde{\bm{\Psi}}(0)=\tilde{\bm{\Psi}}(\omega=0)$. This result is to be compared with the standard Heisenberg operators for the quadratic case, 
	\[e^{iH\Delta t}\bm{\psi}e^{-iH\Delta t}=\exp(-i\Pi K\Delta t)\bm{\psi}
	\]
	and has a clear geometrical meaning: \emph{a rigid translation of the time sites reproduces the conventional time evolution of 
		physical 
		operators}. The details can be found in Appendix \ref{ApD}. This result also holds in the time-dependent case by replacing $e^{i\mathcal{P}_t\Delta t}$ with the unitary ``complete'' time-translation operator $\mathcal{W}(\Delta t)$ from Eq.\ (\ref{eq:wdelta}) which translates both the time sites and the explicit time dependence of $H$ such that $[{\cal W}(\Delta t),{\cal J}]=0$ (see Appendix \ref{ApE}).

	From Eq.\ (\ref{eq:quadraticev}) it follows that if $\mathcal{O}(t)=e^{i\mathcal{P}_t \Delta t}O(\tilde{\bm{\Psi}}(0))e^{-i\mathcal{P}_t \Delta t}$ for 
	 $O$  an arbitrary function of  $\tilde{\bm{\Psi}}(0)$,   then 
	\begin{equation}\label{eq:obs}
	\langle \Phi|\mathcal{O}(t)|\Psi\rangle=\langle \varphi|O_H(t)|\psi\rangle
	\end{equation}
	for  $O_H(t)=e^{iH\Delta t}O(\bm{\psi})e^{-iH\Delta t}$ and  $|\Psi\rangle$ ($|\Phi\rangle$) the history of $|\psi\rangle$ ($|\varphi\rangle$), a relation which holds for \emph{any} quadratic Hamiltonian, observable and states. The generalization to the time-dependent case and multiple-time correlation functions is apparent.
	
	Moreover, time translations preserve the separation between the $\omega=0$ mode  and the rest,  implying 
	\begin{equation}\label{eq:propag}
	\frac{\langle \Phi|e^{-i\mathcal{P}_t\Delta t}|\Psi\rangle}{\langle \tilde{\Omega}|e^{-i\mathcal{P}_t\Delta t}|\tilde{\Omega}\rangle}=\frac{\langle \varphi|e^{-i H(\bm{a},\bm{a}^\dag)\Delta t}|\psi\rangle}{\langle 0|e^{-i H(\bm{a},\bm{a}^\dag)\Delta t}|0\rangle}\,,
	\end{equation} 
	which reduces to Eq. (\ref{eq:product}) for $t=t_0$. An explicit derivation of Eq. (\ref{eq:propag}) is provided in the Appendix \ref{ApD}, which also shows its invariance  under linear symmetries of $\mathcal{J}$ (non-necessary diagonal in time). Its time-dependent version is derived in Appendix \ref{ApE}.

	\subsection{Second Quantization of Parameterized Particles and PaW formalism \label{S4D}}
	One important motivation of the present formulation was to remove the asymmetry between ``space'' and time in QM by incorporating the latter in the same framework.
	Different aspects of this problem are treated in the quantization of reparameterization invariant systems \cite{we.20,pam.50} and related quantum formalisms like the one proposed by Page and Wootters \cite{PaW.83} (and recent revisions \cite{QT.15,lm.20,b.18,m.18,sm.19}, including the relativistic extensions \cite{di.19,dia.19} relevant for the present scheme). Here we discuss how these other proposal are connected to our work through the sp space of particular spaces $\mathcal{H}$. 
	
	The treatment of a parameterized particle (one dimensional for simplicity) for a time independent Lagrangian $L(q,\dot{q})$ leads to a classical weak constraint \cite{pam.50} $H_S=p_t+H\approx 0$ with $p_t=\frac{\partial (\dot{t}L(q,\dot{q}/\dot{t}))}{\partial \dot{t}}$. This condition is quantized as \cite{qg, rov.14}
	\begin{equation}\label{eq:superh}
	H_S|\Psi\rangle= (P_t\otimes \mathbbm{1}+\mathbbm{1} \otimes H) |\Psi\rangle=0\,,
	\end{equation}
	where $P_t\otimes \mathbbm{1}=i\int dt dt' dq\,\frac{d}{dt'}\delta (t'-t)|t q\rangle \langle t' q|$, $\mathbbm{1}\otimes H=\int dt dqdq'\,\langle q'|H|q\rangle |t q'\rangle \langle t q|$ and 
	\begin{equation}\label{eq:algsp}
	    \langle t'q'|tq\rangle=\delta(t-t')\delta(q-q')\,,
	\end{equation}
	which is commonly considered as an auxiliary condition on a ``kinematic space'' $\mathcal{K}$ to define the physical space (which is not a proper subspace). Alternatively, a relational interpretation is assigned to this equation where $H_S$ is regarded as the Hamiltonian of a composite global system ``clock''+``system''. This is the case of the PaW formalism where an hermitian time operator is defined as the observable of the clock $T=\int dt\, t|tq\rangle\langle tq|$.
	
	If instead the kinematic space is promoted to the status of a ``physical'' space and, moreover, the particles are regarded as a $d+1$ dimensional objects (for $d$ spatial dimensions), the proper scenario for many identical particles is an extended Fock space $\mathcal{H}$ \cite{dia.19}, different from the conventional one and \emph{different from the PaW formalism applied to a Fock space} (or equivalently, from the generalized Hamiltonian dynamics of a conventional Fock space). This is achieved by reinterpreting the states $|tq\rangle$ as sp states $|tq\rangle= A^\dag(t,q)|\Omega\rangle$ (with $A(t,q)|\Omega\rangle=0, \langle\Omega|\Omega\rangle=1$) which, considering Eq.\ (\ref{eq:algsp}) and a bosonic particle, implies $[A(t,q),A^\dag(t',q')]=\delta(t-t')\delta(q-q')$, an example of (\ref{eq:alg}).
	Then one may generalize
	\begin{equation}
	    H_S\to -\mathcal{J}
	\end{equation}
	 with
	\begin{equation}\label{eq:jfield}
	\mathcal{J}=\!\int\!dt\! \int\!dq dq' A^\dag(t,q') [i\partial_t\delta(q-q')-\langle q'|H|q\rangle]A(t,q)
	\end{equation}
	which remarkably is the \emph{space-time quantum action} (\ref{eq:action}) for a field of harmonic oscillators (here $i\to q$) and a single particle Hamiltonian (for a local $H$, $\mathcal{J}$ becomes \emph{local in space-time}), a particular instance of the general quadratic case (\ref{eq:quadraticj}). 
	As a consequence, sp states (but not multiparticle states) in $\mathcal{H}$ are formally identical to PaW states while the sp matrix elements of the operators $\mathcal{J}, \mathcal{T}$ are equal to the matrix elements of $H_S, T$ respectively (including $\mathcal{J}|\Psi\rangle=0$ for $|\Psi\rangle\in \mathcal{H}_p$ being formally equivalent to Eq.\ (\ref{eq:superh}) for sp states).
	Notice however that the product structure between ``time'' and ``rest'', essential for ``conditioning on a clock'', is completely lost \cite{dia.19}: The product structure of $\mathcal{H}$ is applied to time itself with a geometrical rather than relational meaning. As a consequence, our definition of foliation (of Sec.\ \ref{S4D}) works on a different basis without any reference to a clock. 
	
	\begin{figure}[h!]
		\centering
		\includegraphics[width=0.42\textwidth]{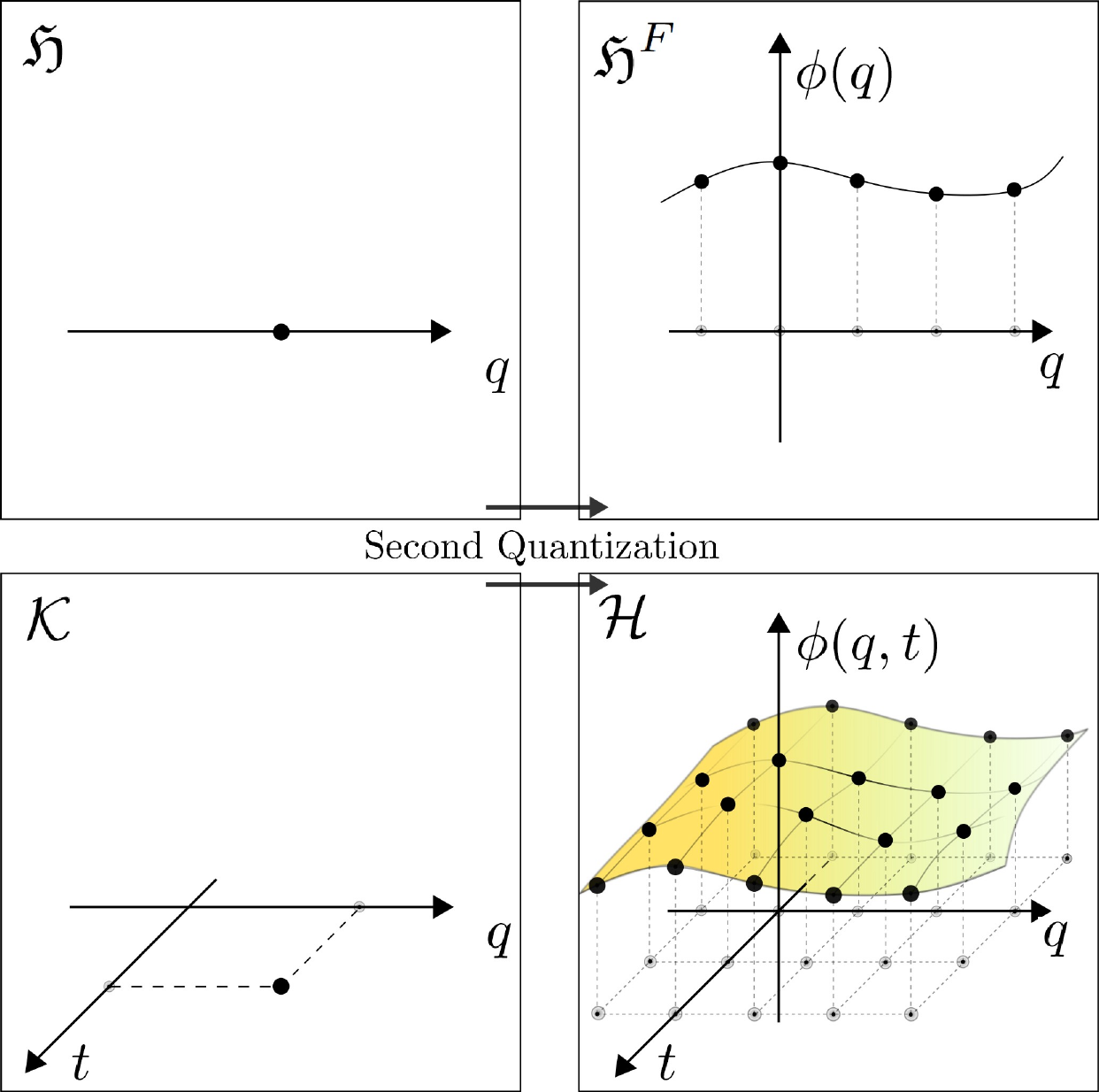}
		\caption{On the left, the two descriptions of the single particle: The conventional one in the Hilbert space  $\mathfrak{H}$ (top panel) and the generalized description in space-time in the Hilbert $\mathcal{K}$ (bottom panel). On the right, the second quantization of the previous schemes. The second quantization of $\mathfrak{H}$ leads to a field theory in a conventional Hilbert space  $\mathfrak{H}^F$ which is isomorphic to a tensor product in space of copies of $\mathfrak{H}$, i.e. $\mathfrak{H}^F\approx \bigotimes_q \mathfrak{H}_q$ (top right panel). The second quantization of $\mathcal{K}$ leads instead to an extended space $\mathcal{H}\approx\bigotimes_t \mathfrak{H}^F_t=\bigotimes_{t,q}\mathfrak{H}_{tq}$ where the tensor product structure is applied to both space and time and it is possible to represent field configurations in space-time (bottom right panel). The description of the field in this extended Hilbert space can be immediately obtained by applying the formalism presented in this work to this particular case.}
		\label{fig:4}
	\end{figure}
	Note also that the second quantization \footnote{By ``second quantization'', we indicate the rigorous mathematical scheme which, given a certain building block (in the present case the states $|tq\rangle$ and $|q\rangle$), allow to construct a Fock space of these indistinguishable elements. See e.g. \cite{Sc.97}. Not to be confused with the historical arguments of QFT \cite{wei.95} .} of the conventional Hilbert space $\mathfrak{H}$ of the particle, which is spanned by states $|q\rangle$, leads as well to a field theory, now in a Fock space $\mathfrak{H}^F$ generated by operators $a^\dag(q)$ such that $|q\rangle=a^\dag(q)|0\rangle$. This is the system described in the present Hilbert $\mathcal{H}$: $\mathcal{J}$ in Eq.\ (\ref{eq:jfield}) is precisely the space-time quantum action which corresponds to the Hamiltonian 
	\begin{equation}
	    H=\int dqdq'\langle q'|H|q\rangle a^\dag(q')a(q)
	\end{equation}
	obtained through second quantization of the Hamiltonian of the particle. 
	The relation between these different Hilbert spaces is represented in Fig.\ \ref{fig:4}. An independent description of the particle (without the field) can be provided in a different $\mathcal{H}$ for $H$ the Hamiltonian of the particle in Eq.\ (\ref{eq:action}).
	
	We remark finally, that while in $\mathfrak{H}^F$ the product structure applied to space allows to represent field configurations at a given time as eigenstates \cite{Note1} 
	\begin{equation}|\phi(q)\rangle=\exp[-\tfrac{1}{2}\int dq\, [a^\dag(q)(a^\dag(q)-2\sqrt{2}\phi(q)) ]|0\rangle\end{equation} of $\phi(q)=\frac{a(q)+a^\dag(q)}{\sqrt{2}}$, in $\mathcal{H}$ the product structure is extended to time allowing to represent space-time configurations \begin{equation}\label{eq:phiqt}
	|\phi(q,t)\rangle=\exp[-\tfrac{1}{2}\!\!\int \!\!dtdq [A^\dag(t,q)(A^\dag(t,q)-2\sqrt{2}\phi(t,q)) ]|\Omega\rangle\,,\end{equation} i.e. Eq.\ (\ref{eq:qt}) applied to the present case.

	\subsection{Relativistic Considerations}
	The relativistic case was traditionally considered as a special case of non-relativistic QM \cite{dy.94} since, e.g. scalar field theories can be interpreted as the continuum limit of coupled harmonic oscillators in space, an example of (\ref{eq:quadraticj}) for free theories. On other hand, the present formalism is particularly suited for a geometrical interpretation of the space-time sites: For $i\to \textbf{x}$ and $\bm{A}_i(t)\to A(x)$, we define $U(\Lambda)$ 
	by $U^\dag(\Lambda)A(x)U(\Lambda)=A(\Lambda x)$ (for $\T\to \infty$). The algebra implied by  Eq.\ (\ref{eq:alg}),  
	\begin{equation} [A(x),A^\dag(y)]=\delta^{(4)}(x-y)
	\end{equation}
	is explicitly preserved when $\Lambda$ is a Lorentz transformation. This yields $U(\Lambda)|\phi(x)\rangle=|\phi(\Lambda^{-1}x)\rangle$ for the coherent field state 
	\begin{equation}|\phi(x)\rangle=\exp[\int d^4x\,\phi(x)A^\dag(x)]|\Omega\rangle\,\end{equation}
	($\alpha(t)\rightarrow\phi(x)$ in  (\ref{eq:coherentst})), which is the correct transformation property 
	of a \emph{state representing a (scalar) field configuration in space-time} (a similar reasoning holds for the states (\ref{eq:phiqt}) for $q\to \textbf{x}$). 
	
	The generator of time translations transforms as $U^\dag(\Lambda)\mathcal{P}_tU(\Lambda)=\Lambda_0^{\;\;\mu}\mathcal{P}_{\mu}$ with $\mathcal{P}_{\mu}:=\int d^4x\, A^\dag(x)i\partial_\mu A(x)$ such that $\mathcal{P}_0=\mathcal{P}_t$. In particular,  $[U(\Lambda),\mathcal{P}_t]=0$ only in the limit of \emph{Galilean transformations}.
	In order to introduce invariant physical subspaces
	we can employ a previous proposal by the authors \cite{dia.19} (more recently also presented in \cite{ve.20})
	which consists of considering a second quantization version of the constraint
	$H^{\text{rel}}_s|\Psi\rangle:=(P^\mu P_\mu -m_0^2)|\Psi\rangle=0$
	(and $P^0>0$) where the hermitian operators $P^\mu$ satisfy $[X^\mu,P_\nu]=i\delta^{\mu}_{\;\nu}$ with $X^0=T$ the PaW time operator \cite{dia.19}. The constraint $H_s^{\text{rel}}|\Psi\rangle=0$ also arises from the treatment of reparameterization invariant systems but considering now the classical action $S=-m_0\int d\tau$ \cite{qg,rov.14}. 
	This treatment leads to 
	\begin{equation}
	H_S^{\text{rel}}\to \mathcal{J}_{\text{rel}}
	=-\int d^4x\,A^\dag(x)(\partial^2+m_0^2)A(x)\,
	\end{equation}
	such that $[U(\Lambda),\mathcal{J}_{\text{rel}}]=0$ and implying
	\begin{equation}\label{eq:relaction}
	\frac{\langle \phi(x)|\mathcal{J}_{\text{rel}}|\phi(x)\rangle}{\langle \phi(x)|\phi(x)\rangle}=S[\phi(x),\phi^\ast(x)]\,
	\end{equation}
	where $S[\phi(x),\phi^\ast(x)]=-\int d^4x\, \phi^\ast(x)(\partial^2+m_0^2)\phi(x)$ is the classical \emph{action of a free scalar field} ($\eta_{00}=1$, $c=1$).
	The result (\ref{eq:relaction}) is suggesting
	a deep connection 
	between particle-like techniques 
	and a formulation of QFT in this extended setting.
	
	This new form of the quantum action also admits a normal decomposition (analogous to (\ref{eq:josc})) such that $[\mathcal{J}_{\text{rel}},A^\dag(m^2,\textbf{p})]=(m^2-m_0^2)A^\dag(m^2,\textbf{p})$ implying in each mass sector the three-dimensional invariant  product \cite{dia.19}. 
	As a consequence, the correct commutators between physical field operators (the component of $\phi(x)\propto A(x)+A^\dag(x)$ at fixed mass) also emerge \cite{dietal2021}. In fact, the definition (\ref{eq:physsubs}) of physical states corresponds in this case  to the mass-shell condition (see also \cite{dia.19}).
	
	Note that we could have considered instead $\mathcal{J}=\int d^4p\, (p^0-E_{pm})A^\dag(p)A(p)$ which yields an equivalent constraint for $E_{pm}=\sqrt{p^2+m^2}$. This $\mathcal{J}$ has the form (\ref{eq:action}) for $H=\int d^3p\, E_{pm}a^\dag(\textbf{p})a(\textbf{p})$ with $[a(\textbf{p}),a^\dag(\textbf{p}')]=\delta^{(3)}(\textbf{p}-\textbf{p}')$, i.e. $H$ is the (diagonalized and normal-ordered) Hamiltonian of the free scalar field we want to describe. While explicit Lorentz symmetry is lost, under e.g. a boost in the first direction such that $p^0\to \cosh{\eta}p^0+\sinh{\eta}p^1$, $U^\dag(\Lambda)\mathcal{J}U(\Lambda)=\cosh{\eta}\mathcal{J}$ and the physical subspace remains invariant: \[[\mathcal{J},\tilde{A}^\dag]=0\Leftrightarrow [U^\dag(\Lambda)\mathcal{J}U(\Lambda),\tilde{A}^\dag]=0 \,.\]
	
	We see that the possibility to represent space-time configurations of the fields opens the possibility to explicitly preserve the symmetries of space-time (Lorentz covariance in the previous example) at the Hilbert space level and in particular in quantization processes. As a fundamental consequence, the correct invariant product emerge in $\mathcal{H}_p$ from the (standard) global inner product of $\mathcal{H}$ in the case considered \cite{dia.19}. 
	
		\section{Recovering physical predictions in the general case \label{IV}}\label{sec:prop}
	\subsection{Quantum Foliations}
	For nonquadratic theories Eq.\ \eqref{eq:quadraticev} (and its time-dependent version) no longer holds for ${\cal V}$ diagonal in time as defined in Eq.\ \eqref{eq:unitary}. 
	However, even for such diagonal solutions, there is still a simple 
	scheme to extract information ``at a given time'' from $|\Psi\rangle$: We introduce a \emph{unitary quantum foliation operator} defined as the shifted inverse FT  	$\tilde{\mathcal{F}}^{\dag}(t) \bm{\tilde{A}}(\omega)\tilde{\mathcal{F}}(t):=\sqrt{\epsilon}\bm{\tilde{A}}(t+\epsilon \T \omega/2\pi)$ such that, roughly speaking, $\tilde{\mathcal{F}}^\dag(t)|\Psi\rangle$ contains the state $U(t,t_0)|\psi\rangle$ at the site $t$.  
 We can make this statement more precise for discrete time in which case (see Appendix \ref{ApA})
	\begin{equation}\label{eq:foliation}
	\tilde{{\cal F}}^\dag(t_j)\tilde{\bm A}(\omega_k)\tilde{{\cal F}}(t_j)=\tilde{\bm A}_{t_{j+k}}\,,
	\end{equation}  implying  \begin{eqnarray}\tilde{\mathcal{F}}^\dag(t)\prod_i [(\tilde{A}^\dag_i(\omega=0))^{n_i}]|\tilde{\Omega}\rangle&=&\prod_i[(\tilde{A}^\dag_{it})^{n_i}]|\tilde{\Omega}\rangle\nonumber\\&=&\mathcal{V}^\dag \prod_{i}[(A^\dag_{it})^{n_i}]|\Omega\rangle\;\;\;\end{eqnarray}
	when $t=t_j=\epsilon j$.
	Hence, given $|\Psi\rangle={\cal L}(|\psi\rangle)\in {\cal H}_P$, 
	we obtain
	\begin{equation}
	\tilde{\mathcal{F}}^\dag(t)|\Psi\rangle= |\psi(t)\rangle_{j}\bigotimes_{j'\neq j}|0(t_{j'})\rangle_{t_{j'}}\label{20}
	\end{equation}
	for $|\psi(t)\rangle=U(t,t_0)|\psi\rangle$, $|0(t)\rangle=U(t,t_0)|0\rangle$ and where we used $\mathcal{V}^\dag=\bigotimes_t U(t,t_0)$. 
	The unitarity of $\tilde{\mathcal{F}}(t)$ reflects the unitarity of time evolution:
	\begin{equation}\label{eq:funit}
	\langle \Phi| \tilde{\mathcal{F}}^\dag(t)\tilde{\mathcal{F}}(t)|\Psi\rangle=\langle \Phi|\Psi\rangle=\langle \varphi|\psi\rangle\;\forall t\,
	\end{equation}
	for $|\Phi\rangle=\mathcal{L}(|\varphi\rangle)$ and in agreement with (\ref{eq:product}).
	
	We see that we can recover the evolved state $|\psi(t)\rangle$ from $|\Psi\rangle$ by first applying the foliation operator and then taking the partial trace over the Hilbert spaces of the other times. This defines a CPTP (completely positive trace preserving) map \cite{nie.01} which in particular for $t=t_0$ provides a representation of $\mathcal{L}^{-1}$. 
	On the other hand, there are straightforward ways to obtain physical predictions which employ the \emph{inner product of the global space} $\mathcal{H}$. In the following we present results in this direction.

	\subsubsection{Propagators}
	Consider again $|\Psi\rangle=\mathcal{L}(|\psi\rangle)$ and $|\Phi\rangle=\mathcal{L}(|\varphi\rangle)$.
	From Eq.\ \eqref{20} it follows that
	\begin{equation}\label{eq:prop}
	\begin{split}
	\frac{\langle \Phi|\tilde{\mathcal{F}}(t_0)e^{-i\mathcal{P}_t(t-t_0)}\tilde{\mathcal{F}}^\dag(t)|\Psi\rangle}{\langle \tilde{\Omega}|e^{-i\mathcal{P}_t(t-t_0)}|\tilde{\Omega}\rangle}=\frac{\langle \varphi|U(t,t_0)|\psi\rangle}{\langle 0|U(t,t_0)|0\rangle}
	\end{split}
	\end{equation}
	with $\langle \varphi|U(t,t_0)|\psi\rangle$ \emph{the standard propagator}. Here $e^{-i{\cal P}_t(t-t_0)}$ moves $|\psi(t)\rangle$ (and the remaining vacua) back to site $t_0$ where it overlaps $\langle \varphi|$. 
	The remaining overlaps between vacua cancel with those in the denominator.  For $t=t_0$  (\ref{eq:funit}) is recovered.

	The result (\ref{eq:prop}) can be easily written in terms of the original operators ${\bm A}(\omega=0)$, ${\bm A}^\dagger(\omega=0)$ or $\bm{A}(t)$,   $\bm{A}^\dag(t)$. 
	For time-independent $H$,  where time translations are a symmetry ($[\mathcal{P}_t,\mathcal{J}]=0$) the following simple expressions can be obtained ($\Delta t=t-t_0$): 
	    \begin{align}
	\frac{\langle \Phi|\tilde{\mathcal{F}}(t_0)e^{-i\mathcal{P}_t\Delta t}\tilde{\mathcal{F}}^\dag(t)|\Psi\rangle}{\langle \tilde{\Omega}|e^{-i\mathcal{P}_t\Delta t}|\tilde{\Omega}\rangle}=&\,\frac{{}_0\langle \Phi|e^{-i H(\bm{A}(0),\bm{A}^\dag(0))\Delta t}|\Psi\rangle_0}{\langle \Omega| e^{-i H(\bm{A}(0),\bm{A}^\dag(0))\Delta t}|\Omega\rangle}
	\label{eq:propag2}
	\\=&\,\frac{{}_0\langle \Phi|\mathcal{F}(t)e^{i\mathcal{J}\Delta t}\mathcal{F}^\dag(t_0)|\Psi\rangle_0}{\langle \Omega|e^{i\mathcal{J}\Delta t}|\Omega\rangle}\label{eq:propag3}\,,
	\end{align}
	where $|\Psi\rangle_0$, $|\Phi\rangle_0$, ${\cal F}(t)$ 
	are in the trivial basis
	(see Appendix \ref{ApF} for the proof). Clearly, Eq.\ (\ref{eq:propag2}) agrees with Eq.\ (\ref{eq:prop}) and its limit $\epsilon\to 0^+$ is well defined. 
	In the quadratic case, this equation reduces to \eqref{eq:propag} since $[{\cal F}(t),\int dt \,{\cal H}(\bm{A}(t),\bm{A}^\dag(t))]=0$ 
	for ${\cal H}(\bm{A}(t),\bm{A}^\dag(t))=\sum_i\omega^i_0A^\dag_i(t){A}_i(t)$.  
	The generalization for a time-dependent $H$
	relies on the replacement $e^{i\mathcal{P}_t(t-t_0)}\to \mathcal{W}(t-t_0)$ and is developed in the Appendix \ref{ApE}.
    
	\subsubsection{Observables and Correlation Functions}
\begin{figure}[ht]
		\centering
		\includegraphics[width=0.482\textwidth]{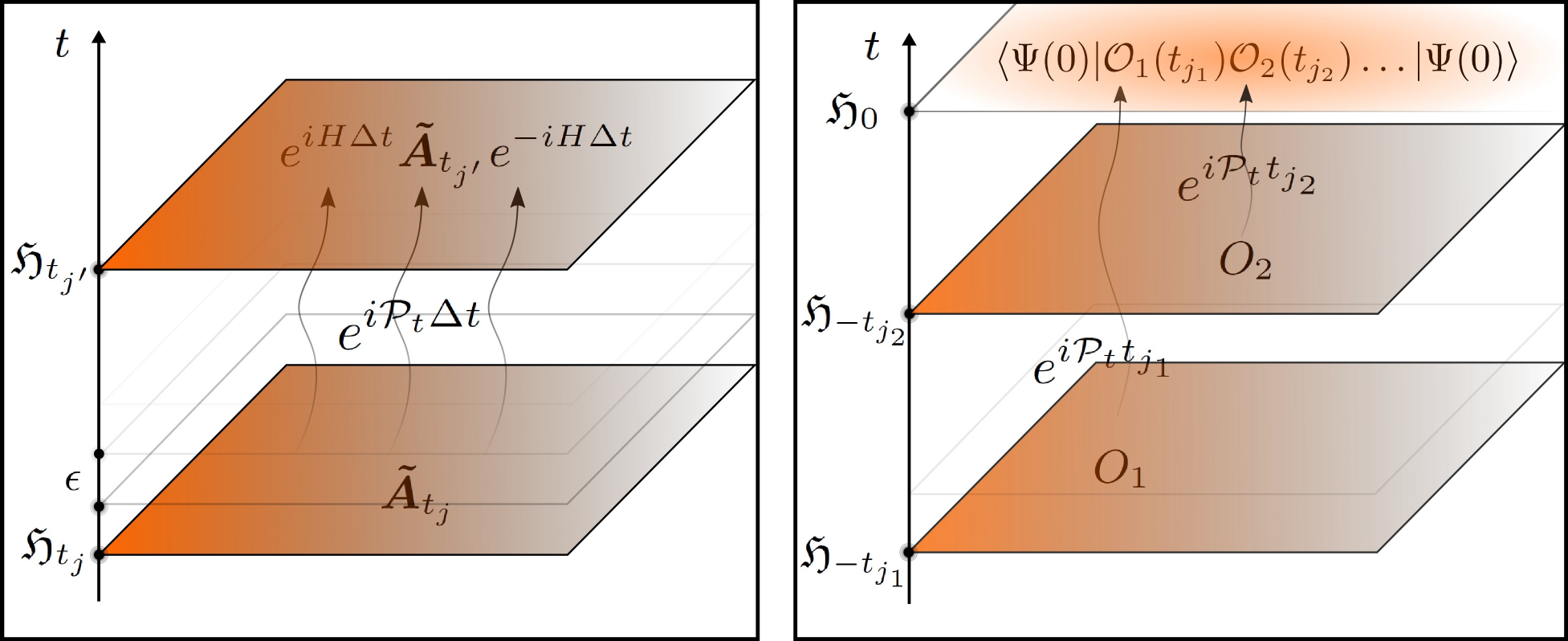}
		\caption{
			Under time translations through $\Delta t/\epsilon$ steps, the operator $\bm{\tilde{A}}_{t_j}$ is displaced to site $t_{j'}=t_j+\Delta t$ while evolving an amount $\Delta t$ (left panel). Through  insertion of operators at different times and translations back to the Hilbert at $t=0$ multiple-time correlation functions are obtained (right panel).}
		\label{fig:3}
	\end{figure}

	For $H$ time-independent, Eq.\ (\ref{eq:Jpt}) allows us to write (see also Eq.\ (\ref{eq:finitep}))
	\begin{eqnarray}\label{eq:tildetransl}
	e^{i\mathcal{P}_t\epsilon}\bm{\tilde{A}}_{t_j}e^{-i\mathcal{P}_t\epsilon}
	&=&e^{iH\epsilon}\bm{\tilde{A}}_{t_{j+1}}e^{-iH\epsilon}
	\end{eqnarray}
	with $H$ the Hamiltonian as a function of operators $\bm{\tilde{A}}_{t_{i+1}}$, $\bm{\tilde{A}}^\dag_{t_{i+1}}$. We see that under the action of time translations, the operators $\bm{\tilde{A}}_{t_i}$ not only are translated into the new Hilbert, but they are also evolving (see Figure \ref{fig:3}). More generally,  (\ref{eq:tildetransl}) implies
	\begin{equation}\label{eq:id58}
	e^{i\mathcal{P}_t\Delta t}O(\bm{\tilde{A}}_{t_j},\bm{\tilde{A}}^\dag_{t_j})e^{-i\mathcal{P}_t\Delta t}=e^{iH \Delta t}O(\bm{\tilde{A}}_{t_{j'}},\bm{\tilde{A}}^\dag_{t_{j'}})e^{-iH\Delta t}\,
	\end{equation}
	with $H\equiv H(\bm{\tilde{A}}_{t_{j'}},\bm{\tilde{A}}^\dag_{t_{j'}})$ and $\Delta t=t_{j'}-t_j$.
	
	We can employ this point of view to obtain correlation functions: Given a conventional operator $O(\bm{a},\bm{a}^\dag)$, which in the Heisenberg picture reads $O_H(t)=e^{i H t}Oe^{-i H t}$ (we set $t_0=0$), from  \eqref{eq:id58} we obtain 
	\begin{eqnarray}\label{eq:corr}
	\langle \varphi|O_H(t_j)|\psi\rangle&=&\langle \Phi(0)|e^{i\mathcal{P}_tt_j}O(\tilde{\bm A}_{-t_j},\tilde{\bm A}^\dag_{-t_j}) e^{-i\mathcal{P}_t t_j}|\Psi(0)\rangle
	\nonumber\\
	&=&\langle \Phi|e^{iHt_j}O(\bm{\tilde{A}}(0),\bm{\tilde{A}}^\dag(0))e^{-iHt_j}|\Psi\rangle\,,
	\end{eqnarray}
	for $|\Psi(0)\rangle=\tilde{\mathcal{F}}^\dag(0)|\Psi\rangle$, $|\Psi\rangle=\mathcal{L}(|\psi\rangle)$, $|\Phi(0)\rangle=\tilde{\mathcal{F}}^\dag(0)|\Phi\rangle$, $|\Phi\rangle=\mathcal{L}(|\varphi\rangle)$.
	In the last equality we have ``extracted'' the operators $\tilde{\mathcal{F}}(0)$ from $|\Psi(0)\rangle$, $|\Phi(0)\rangle$, 
	such that $\bm{\tilde{A}}(0)=\bm{\tilde{A}}(\omega=0)$ and $H\equiv H(\bm{\tilde{A}}(0),\bm{\tilde{A}}^\dag(0))$. This is of course the expression which is obtained by applying the map $\mathcal{L}$ to both the states $|\psi\rangle$, $|\varphi\rangle$ and the operator $O$. 
	
	The result (\ref{eq:corr}) can be immediately generalized to compute multiple-time correlation functions by ``inserting'' now operators at different times:
	If we define \[\mathcal{O}^i(t_{j_i}):=e^{i\mathcal{P}_t t_{j_i}}O^i(\bm{\tilde{A}}_{-t_{j_i}},\bm{\tilde{A}}^\dag_{-t_{j_i}})e^{-i\mathcal{P}_t t_{j_i}}\,\]
	then
    \begin{equation}\label{eq:corr2}
    \langle \varphi|\prod_i O^i_H(t_{j_i}) |\psi\rangle=\langle \Phi(0)|\prod_i \mathcal{O}^i(t_{j_i})|\Psi(0)\rangle\,.
	\end{equation}
	The corresponding $\omega$ expansion is apparent and only involves physical operators (operators acting on $\mathcal{H}_p$).
	
	All these relations, starting from Eq.\ (\ref{eq:tildetransl}), can be generalized to the time-dependent case by replacing $e^{i\mathcal{P}_t \Delta t}\to \mathcal{W}(\Delta t)$ from Appendix \ref{ApE}. 
	A similar procedure can be employed for the mixed case and for the more general decoherence functional \cite{gellm.19}.

	\subsection{Path Integrals from Quantum Trajectories}\label{S:PI}
The space-time quantum actions, their unitary equivalence with $\mathcal{P}_t$ and the ``trajectory'' states (\ref{eq:coherentst})-(\ref{eq:qpt}) 
also enable a straightforward
novel approach to path integrals (PIs), which provides an alternative way to compute 
physical predictions. In order to illustrate this point, we will show first that a conventional product of time ordered operators in $\mathfrak{H}$ can be expressed in ${\cal H}$ as
	   \begin{equation}\label{eq:iden}
	       \hat{T}\left( O^1_H(t_1)O^2_H(t_2)\dots O^n_H(t_n)\right)={\rm Tr}_{t\neq 0}\left[e^{i\mathcal{J}\epsilon} \mathcal{O}\right]\,, 
	   \end{equation}
where $O^i_H(t)=U^\dag(t,0)O^iU(t,0)$, $t_i=\epsilon j_i$, and 
	\begin{equation}
\mathcal{O}:=O^1(\bm{A}_{t_{j_1}},\bm{A}^\dag_{t_{j_1}}) \dots O^n(\bm{A}_{t_{j_n}},\bm{A}^\dag_{t_{j_n}}) \label{eq:Ocal}
	\end{equation}
	is a product operator in time with $O^i$ on the slice $\mathfrak{H}_{t_{j_i}}$ (and identities for $j\neq j_i$). The time-ordering emerges naturally from the ordering of the time-sites in $\mathcal{H}$. 
	This also provides an alternative representation of the product of operators in (\ref{eq:corr2}) (when the times are ordered): $\prod_i {\mathcal O}^i(t_{j_i})=
	{\rm Tr}_{t\neq 0}\left[e^{i\mathcal{J}\epsilon} \mathcal{O}\right]$.
	
{\it Proof}: Note first that Eq.\ (\ref{eq:iden}) is equivalent to $\langle \varphi|  \hat{T} O^1_H(t_1)O^2_H(t_2)\dots O^n_H(t_n)|\psi\rangle={\rm Tr}\left[|\psi\rangle \langle\varphi|e^{i\mathcal{J}\epsilon} \mathcal{O}\right]$ $\forall$ $|\psi\rangle\langle \varphi
	    |\equiv|\psi\rangle\langle \varphi|\bigotimes_{j\neq0} \mathbbm{1}_j$
	   acting on $\mathfrak{H}_0$. On the other hand, from the result (\ref{eq:jdiag}), and the initial condition $\mathcal{V}^\dag \bm{A}_{t=0}\mathcal{V}=\bm{A}_{t=0}$ (implying $[{\cal V},|\psi\rangle\langle\varphi|]=0$),
	   \begin{equation}\label{eq:trjp}
	       {\rm Tr}[|\psi\rangle \langle\varphi|e^{i\mathcal{J}\epsilon} \mathcal{O}]={\rm Tr}[|\psi\rangle \langle\varphi|e^{i\mathcal{P}_t\epsilon}\mathcal{V}\mathcal{O}\mathcal{V}^\dag]
	   \end{equation}
with \begin{equation}\label{eq:evolvedop}
	    \mathcal{V}\mathcal{O}\mathcal{V}^\dag=O^1_H(\bm{A}_{t_1},\bm{A}^\dag_{t_1},t_1)\dots  O^n_H(\bm{A}_{t_n},\bm{A}^\dag_{t_n},t_n)\,,
	\end{equation}
	i.e. $\mathcal{V}\mathcal{O}\mathcal{V}^\dag$ is a tensor product of operators $O^i_H(t)$, each one evolved up to the corresponding time site value. 
We then  note that a quantity $\langle \varphi|O^1O^2\dots O^n|\psi\rangle$ can be 
rewritten as ($\bm{i}=\dots,i_{-2},i_{-1},i_1,i_2,\dots$)
	\begin{eqnarray}\label{eq:traza}
	\langle \varphi|O^1O^2\dots O^n|\psi\rangle&=&\sum_{\bm{i}}\langle  \varphi i_{1} i_2 \dots|e^{i\mathcal{P}_t\epsilon}\mathcal{O}|\psi i_{1} i_2 \dots\rangle
	\nonumber\\&=&{\rm Tr}\left[|\psi\rangle\langle \varphi|e^{i\mathcal{P}_t\epsilon}\mathcal{O}\right]
	\end{eqnarray}
	with $\sum_i |i\rangle \langle i|=\mathbbm{1}$ 
	and the operators appearing in the inverse order of the time-sites (here, to comply with the ordering on the left hand side of (\ref{eq:traza}) we should choose
	$t_1>t_2>\dots >t_n$ in the definition of $\mathcal{O}$. With the time-ordering operator this is no longer required.). 
 The expression (\ref{eq:traza}) relies on a basic relation between quadratic forms and tensors \cite{ish.94} 
	(e.g.\ $\langle\varphi|O^1 O^2|\psi\rangle=\sum_i \langle\varphi|O^1|i\rangle\langle i|O^2|\psi\rangle=\sum_i\langle \varphi i|e^{i\cal P}O^2\otimes O^1|\psi  i\rangle$ for 
	$e^{-i\cal P}|\varphi i\rangle=|i\varphi\rangle$). 
	In \eqref{eq:traza} the time translation operator $e^{i\mathcal{P}_t\epsilon}$ ensures the correct indices ordering. 
	
	The validity of Eqs.\ \eqref{eq:trjp}-\eqref{eq:traza} $\forall$ $|\psi\rangle\langle\varphi|$ implies \eqref{eq:iden}, with the time-ordering linked to the underlying ordering of the time-sites. \qed
	
	    Now, by using Eq.\ (\ref{eq:iden}) and considering for simplicity (and ease of notation) a Hilbert $\mathfrak{H}$ such that $\int dq\, |q\rangle\langle q|=\mathbbm{1}$,  $|\psi\rangle=|q_i\rangle$, $|\varphi\rangle=|q_f\rangle$
	    and $O^i(a,a^\dag)\equiv O^i(q)$ we can write 
	     \begin{equation}\label{eq:PI}
	        \begin{split}
	           & \langle q_f|\hat{T}O^1_H(q,t_1)\dots O_H^n(q,t_n)|q_i\rangle
	            \\=&\int \prod_{j\neq 0} dq_j\,O^1(q_{j_1})\dots O^n(q_{j_n})
	             \, \langle q_fq_1\dots|e^{i\mathcal{J}\epsilon}| q_iq_1\dots\rangle\,,
	        \end{split}
	    \end{equation}
	    where we used the resolution of the identity in $\mathcal{H}$, $\int \prod_j dq_j |\textbf{q}\rangle\langle \textbf{q}|=\mathbbm{1}$ (here $|\textbf{q}\rangle=\bigotimes_j|q_j\rangle_{t_j}$ satisfying (\ref{eq:qpt})).
	    The right-hand side is formally identical to 
	    the standard PI expansion of this quantity for a periodic evolution (such that $U^\dag(t,0)=U(\T,t)$):
	    \begin{equation*}
	        \begin{split}
	          \langle &q_f|\hat{T}O^1_H(q,t_1)\dots O_H^n(q,t_n)|q_i\rangle\\= &\int \prod_{j\neq 0} \Big[\frac{dq_j}{
	       \sqrt{\frac{2\pi i \epsilon}{m}}}\Big]\frac{1}{\textstyle\sqrt{\frac{2\pi i \epsilon}{m}}}\,O^1(q_{t_1})\dots O^n(q_{t_n})
	             \, e^{iS}\,,
	        \end{split}
	    \end{equation*}
	    with $S$ the classical action for $H(p,q,t)=p^2/2m+V(q,t)$ (not required in Eq. (\ref{eq:PI})) evaluated on each path. 
	    Remarkably, the quantity $e^{iS}$ is now appearing from the matrix elements of $e^{i\mathcal{J}\epsilon}$ along the ``quantum trajectories'' defined by the extended Hilbert space and represented in Fig.\ \ref{fig:1}. This can be seen explicitly by writing first $\langle \textbf{q}|e^{i\mathcal{J}\epsilon}|\textbf{q}\rangle=\int \prod_j dp_j \langle \textbf{q}|\textbf{p}\rangle \langle \textbf{p}|e^{i\mathcal{J}\epsilon}|\textbf{q}\rangle$, 
	   ($|\textbf{p}\rangle=\bigotimes_j|p_j\rangle_{t_j}$ satisfying (\ref{eq:qpt})) and noting that ($t\equiv j$)
	  	\begin{equation}\label{eq:jelements}
	    \langle \textbf{p}|e^{i\mathcal{J}\epsilon}|\textbf{q}\rangle= e^{i\sum_t \epsilon[p_t\dot{q}_t-H(p_t,q_t,t)]}\langle \bm{p}|\bm{q}\rangle+\mathcal{O}(\epsilon^2)\,,
	\end{equation}
	as it follows from the  approximation of the \emph{operator}
	\begin{equation}
	    \begin{split}
	    e^{i\mathcal{J}\epsilon}=& \mathbbm{1}+i\epsilon \mathcal{J}+\mathcal{O}(\epsilon^2)\\=&\mathbbm{1}+i\epsilon\sum_t [p_t\dot{q}_t-H(p_t,q_t,t)]+\mathcal{O}(\epsilon^2)\,, \nonumber
	    \end{split}
	\end{equation} 
	where  $\dot{q}_t=(q_{t+1}-q_t)/\epsilon+\mathcal{O}(\epsilon^2)$, i.e. $\dot{q}_t$ is equal to the site derivative of $q_t$ in this order (see also Eqs.\ (\ref{A3}, \ref{eq:finitep})). 
	    We can corroborate the result (\ref{eq:jelements}) by noting that $e^{i\mathcal{J}\epsilon}=\mathcal{V}^\dag e^{i\mathcal{P}_t\epsilon}\mathcal{V}=e^{i\mathcal{P}_t\epsilon}\bigotimes_j U(t_j+\epsilon,t_j)$ which, by considering again Eq.\ (\ref{eq:traza}) (now ``from right to left''), implies $\langle \textbf{p}|e^{i\mathcal{J}\epsilon}|\textbf{q} \rangle=\prod_j \langle p_j|U(t_j+\epsilon,t_j)|q_{j-1}\rangle$, the expression which is obtained through the conventional time-slicing for a spacing $\epsilon$, in agreement with  (\ref{eq:jelements}). 
	    
	    Related expressions can be derived  for propagators by similar means. Coherent states (\ref{eq:coherentst}) may also be employed analogously. Since the ``sum over trajectories'' interpretation \cite{feyn.05} acquire in $\mathcal{H}$ Hilbert space meaning, the conventional subtleties of PIs concerning the limits $\epsilon\to 0$, $\T \to \infty$ appear now linked to standard issues related to tensor products in Hilbert space. The results and regularizations employed in this work may thus constitute a first step towards tackling those subtleties by means of well-known techniques from canonical QM. 
	    
	   	\section{Discussion\label{V}}
	The treatment presented in this paper
	provides a starting point for developing general space-time formulations. While it is able to reproduce the conventional predictions of QM concerning time evolution, it maps the evolution to history states endowed with a rich time structure. This natural consequence of the underlying product structure in time of the extended Hilbert space opens some immediate possibilities concerning the understanding of time correlations. 
	In particular, such time structure could be relevant in the investigation of the entanglement/geometry connection \cite{van.10,sw.12,ni.18} since it may enable space-time extensions of recent proposals of emerging space from entanglement in Hilbert space \cite{Ca.17}. More generally, quantum correlations across time-like (causally connected) intervals acquire meaning.
	
	In this work, almost all the efforts concerning physical predictions have been focused on the recovering of the conventional consequences of QM. However, the unitary equivalence between theories revealed by the formalism opens additional unexplored possibilities. For example, since all theories are defined in the same Hilbert space $\mathcal{H}$, not only the time evolution of all possible theories follow but, in principle, also that of any quantum superposition of them 
	(being $\mathcal{H}$ a genuine Hilbert space), a situation which may find its place in nature: A coherent superposition of gravity \cite{mar.17,bos.17,mar.20} could induce coherences in the time evolution of matter. A related example is the possibility of introducing indefinite causal-order (superposition of causal relations between events), a problem which also requires a non-trivial extension of QM, recently under consideration in the context of process matrices \cite{cas.18}.
	In these new scenarios non-diagonal in time properties of $\mathcal{P}_t$ and $\mathcal{J}$, some of which have  been discussed in Appendices \ref{ApB} and \ref{ApD}, may become relevant.
	
	While the space-time quantum actions have a natural form for infinite dimensional $\mathfrak{H}$ (they resemble a classical action), 
	the formalism is completely suitable for finite dimensional systems. For instance, since the general evolution of a qubit can be encoded in the first two levels of an harmonic oscillator, ``space-time descriptions'' of a qubit can be derived from subspaces of the present $\mathcal{H}$ (and apparent generalizations to higher dimensions).
	
	The present formulation
	also provides a consistent framework for discretizing time. In Sec.\ \ref{S:PI} this discretization has been related 
	to the conventional time-slicing 
	employed in path integrals through the matrix elements of $\mathcal{J}$. Further developments along this line are under investigation.
	We also mention that in case a fundamental spacing $\epsilon$ exists,
	it would have non-trivial physical implications, as recently shown in \cite{we.20} through the related quantization techniques considered in section \ref{S4D}. 
	Other insights derived from the formulations considered there  may be further  developed in their  ``second quantized version'' as particular instances of this framework. For example, the considerations on Lorentz covariance at the Hilbert space level described in the PaW extensions \cite{di.19,dia.19} for relativistic particles were here generalized to (free) fields. 
	
	While we have employed pure states, the mixed case follows straightforwardly by usual means. Considering in addition that the treatment of composite systems is implicit in (\ref{eq:alg}), the formalism should describe measurements properly by incorporating the processes involved \cite{nie.01},
	a strategy recently employed in related constructions  \cite{QT.15,lm.20}. 
	
	Decoherence functionals can also be derived straightforwardly from the formalism opening
	 possible connections with Isham's approach \cite{ish.93} (and modern related schemes, e.g. \cite{fit.15,cot.18}). 
	In particular, it is interesting that the concept of physical states, which appear naturally in the second quantization of parameterized particles (Sec.\ \ref{S4D}), can be related to such quantity, providing a possible 
	unifying bridge between these different generalizations of QM.

	We note finally that while a bosonic formulation was employed, the formalism is also suited for fermions: Given a set of fermions $b_i$ such that $[b_i,b^\dag_j]_+=\delta_{ij}$, the corresponding operators on each slice $B_i(t), B^\dag_j(t)$ can be defined as
	$
	[B_i(t),B_j^\dag(t')]_+=\delta(t-t')\delta_{ij}\,,
	$
	which in particular implies a \emph{Pauli's exclusion principle in time}. Then, main basic results, starting from the unitary relation (\ref{eq:jdiag}) between $\mathcal{P}_t$ and $\mathcal{J}$ (see Appendix \ref{ApB}), hold if we replace $\bm{A}(t)\to \bm{B}(t)$.
	
	In summary, we presented a formulation of QM which treats time and ``space'' on the same footing at the Hilbert space level. The concept of time evolution is replaced by the notion of physical subspaces determined by new central actors: The space-time quantum actions. All familiar tools of QM can now be applied to this extended framework, paving the way for a novel understanding of quantum correlations across time.

	\acknowledgments
	
	We acknowledge support from CONICET (N.L.D., J.M.M.) and  CIC (R.R.) of Argentina. This work was supported by CONICET PIP Grant No. 11220150100732.

	\appendix
	\section{Regularizations and Notation\label{ApA}}
	In this Appendix we summarize the notation conventions we adopt in relation to the regularizations applied to the extent of time $\T$ and the spacing between sites $\epsilon$. For completeness and clarity, explicit expressions and limits are provided as well.
	
	For finite $\T$, 
	\begin{equation}
	\bm{A}(\omega_k)=\tfrac{1}{\sqrt{T}}\int_{-T/2}^{T/2}dt \bm{A}(t)e^{i\omega_k t}\end{equation}
	with 
	$\omega_k=\frac{2\pi k}{T}$, such that 
	\begin{equation}\bm{A}(t)=\tfrac{1}{\sqrt{\T}}\sum_k \bm{A}(\omega_k)\, e^{-i\omega_k t}\label{A02}\end{equation} with $k\in\mathbb{Z}$ and 
	$[A_i(\omega_k),A^\dag_j(\omega_{k'})]=\delta_{ij}\delta_{kk'}\,.
	$
	By Eq.\ \eqref{A02} the ``site'' derivative \begin{equation}\label{A3}
	    \bm{\dot{A}}(t):=\lim_{\delta t \rightarrow 0} \frac{\bm{A}(t+\delta t)-\bm{A}(t)}{\delta t}\,
	\end{equation}  becomes identical with 	$\tfrac{-i}{\sqrt{\T}}\sum_k \omega_k\bm{A}(\omega_k)\, e^{-i\omega_k t}$. 
	
		In continuous notation, we can rewrite \eqref{A02} as 
	$    \bm{A}(t)=\int \frac{d\omega}{\sqrt{2\pi}} \bm{A}(\omega) e^{-i\omega t}$, 
	where $\bm{A}(\omega)=\sqrt{\frac{\T}{2\pi}}\bm{A}(\omega_k)$ and $\int d\omega$ stands for 
	$\frac{2\pi}{\T}\sum_k $, such that $i\bm{\dot{A}}(t)=\int\frac{d\omega}{\sqrt{2\pi}}\omega\bm{A}(\omega)e^{-i\omega t}$.  For $\T\rightarrow\infty$ this representation becomes exact, with 
	$[A_i(\omega),A^\dag_j(\omega')]\mathop{\longrightarrow}\limits_{\T\rightarrow\infty}\delta_{ij}\delta(\omega-\omega')$. 
	
	On the other hand, for discrete time (finite $\epsilon$), $\bm{A}(\omega_k)$ becomes the discrete FT 
	\begin{equation} \bm{A}(\omega_k)=
	\sqrt{\tfrac{\epsilon}{T}}\sum_j\bm{A}_{t_j} e^{i\omega _k t_j}, \end{equation}
	where $\bm{A}_{t_j}=\sqrt{\epsilon}\bm{A}(t_j)$,  
	$t_j=\epsilon j$, 
	$k,j=-m,\ldots,m$ and $T/\epsilon=2m+1$,  
	with  $[A_{t_j},A_{t_{j'}}]=\delta_{jj'}$, such that 
	 $\bm{A}_{t_j}=\sqrt{\tfrac{\epsilon}{T}}\sum_k\bm{A}(\omega_k) e^{-i\omega _k t_j}$. 
	The last expression can also be used for a continuous $t$,
	in which case $[A_i(t),A^\dag_j(t')]=\delta_{ij}
	\frac{1}{\T}\frac{\sin [\pi (t-t')/\epsilon]}{\sin [\pi(t-t')/\T]}\mathop{\longrightarrow}\limits_{\epsilon\rightarrow 0^+}\delta_{ij}\delta(t-t')$. 
	
	We can also define the one-body unitary operator $\mathcal{F}(t_j)=\exp[-i \bm{A}^\dag(\omega_{k'}) M^{kk'}_{t_j}\bm{A}(\omega_k)]$
	with $(e^{i M_{t_j}})_{kk'}=\sqrt{\frac{\epsilon}{T}}
	e^{-i2\pi (k+j)k'\epsilon/\T}$, such that
	\begin{equation}
	\mathcal{F}^\dag(t_j)\bm{A}(\omega_k)\mathcal{F}(t_j):=\bm{A}_{t_{(j+k)}}\,
	\end{equation}
    with $\mathcal{F}^\dagger(t_j)=e^{i\mathcal{P}_t t_j}\mathcal{F}^\dagger(0)$ and  $\mathcal{F}(0)$ the FT.
 	For finite $\epsilon$, $\mathcal{P}_t$ is still defined as $\mathcal{P}_t=\sum_k \frac{2\pi k}{\T}A^\dag(\omega_k)A(\omega_k)$ where the sum now involves $\T/\epsilon$ values and
	\begin{equation}\label{eq:finitep}
	e^{i\mathcal{P}_t\epsilon}\bm{A}_{t_j} e^{-i\mathcal{P}_t\epsilon}=\bm{A}_{t_{j+1}}\,.
	\end{equation}
	Similarly, for a non-trivial theory, the \emph{physical} foliation operators used in \eqref{eq:foliation} are defined by
	\begin{equation}
	\tilde{\mathcal{F}}^\dagger(t_j)={\cal V}^\dagger {\cal F}^\dagger(t_j) {\cal V}=	e^{ i  {\cal J} t_j}  \tilde{\cal F}^\dagger(0)\,. 
	\end{equation}
	
	\section{Unitary Relation between $\mathcal{P}_t$ and $\mathcal{J}$ and Additional Properties of $\mathcal{P}_t$ \label{ApB}} 
  Here some additional properties of the generator of time translation are presented, starting with the relation between its commutator and the ``partial'' derivative in time. Immediate (but non trivial) consequences follow.
	Before proceeding, an elementary proof of the result (\ref{eq:jdiag}) which only employs Eq.\ (\ref{eq:pt})  is provided below.
	
	\begin{proof}[Proof of Eq.\ \eqref{eq:jdiag}]
For finite $\T$ we assume 
		\begin{equation}
		\hat{T}'\exp[-i \int dt\int_{t_0}^{t_0+\T}dt'\mathscr{H}(\bm{A}(t),\bm{A}^\dag(t),t')]=\mathbbm{1}\,,
		\end{equation}
		i.e. $U(t_0+\T,t_0)=\mathbbm{1}$.
		Then,
		\[
		e^{i\mathcal{P}_t\delta t}\mathcal{V}^\dag e^{-i\mathcal{P}_t\delta t}=\hat{T}'\!\exp[-i\!\int\! dt\!\int_{t_0}^{t-\delta t}\!\!\!\!\! dt'\mathscr{H}(\bm{A}(t),\bm{A}^\dag(t),t')]
		\]
		which holds for $\T\to \infty$ when $\mathscr{H}$ is well-behaved in the limit of large times. For $\delta t\ll 1$, it leads to 
		\[
		e^{i\mathcal{P}_t\delta t}\mathcal{V}^\dag e^{-i\mathcal{P}_t\delta t}=e^{i\delta t \int dt\, \mathscr{H}(\bm{A}(t),\bm{A}^\dag(t),t)}\mathcal{V}^\dag\,,
		\]
		where we used 
		$\int_{t_0}^{t-\delta t} dt'\, \mathscr{H}(\bm{A}(t),\bm{A}^\dag(t),t')\approx\int_{t_0}^{t} dt'\, \mathscr{H}(\bm{A}(t),\bm{A}^\dag(t),t')-\delta t \mathscr{H}(\bm{A}(t),\bm{A}^\dag(t),t)$ and the temporal ordering (the second term is always at time $t>t'$). 
		In conclusion,
		\begin{equation}
		\mathcal{V}^\dag e^{-i\mathcal{P}_t\delta t}\mathcal{V}=e^{-i[\mathcal{P}_t-\int dt\,\mathscr{H}(\bm{A}(t),\bm{A}^\dag(t),t)]\delta t}
		\end{equation}
		implying $\mathcal{V}^\dag \mathcal{P}_t \mathcal{V}=\mathcal{J}$. 
	\end{proof}
	
	Notice that this proof only employs properties of $\mathcal{V}^\dag$ under time translations. In particular, this means that it also holds for \emph{fermionic systems}.
	
	We can actually prove, 
	for a general 
	periodic operator 
	\begin{equation}
	{\cal U}=\exp[\int dt\,{\cal M}(\bm{A}(t),\bm{A}^\dag(t),t)]\,,
	\end{equation}
	the more general result 
	\begin{eqnarray} {\cal U}{\cal P}_t{\cal U}^{-1}&=&{\cal P}_t-i \left(\frac{\partial{\cal U}}{\partial t}\right){\cal U}^{-1}\\&=&{\cal P}_t-i\int  dt\,{\cal R}(\bm{A}(t),\bm{A}^\dag(t),t)\,,\label{A4}
	\end{eqnarray}
	which is equivalent to 
	\begin{equation}[{\cal P}_t,{\cal U}]=i\frac{\partial{\cal U}}{\partial t}\,,\label{A5}
	\end{equation}
	where its partial derivative is defined as  
	\begin{eqnarray}\label{eq:derivative}
	\frac{\partial {\cal U}}{\partial t}:&=&
	\lim_{\delta t\rightarrow 0}
	\frac{e^{\int dt\,{\cal M}(\bm{A}(t),\bm{A}^\dag(t),t+\delta t)}-{\cal U}}{\delta t}\nonumber\\&=&\left(\int dt\, {\cal R}(\bm{A}(t),\bm{A}^\dag(t),t)
	\right){\cal U}
	\end{eqnarray}
	and ${\cal R}(\bm{A}(t),\bm{A}^\dag(t),t)$ is the operator defined by 
	$\frac{\partial}{\partial t'} e^{{\cal M}(\bm{A}(t),\bm{A}^\dag(t),t')}
	={\cal R}(\bm{A}(t),\bm{A}^\dag(t),t')
	e^{{\cal M}(\bm{A}(t),\bm{A}^\dag(t),t')}$. 
	\begin{proof}
		Using previous definitions we obtain,  
		up to $O(\delta t)$, 
		\begin{eqnarray}e^{i\mathcal{P}_t\delta t}\mathcal{U} e^{-i\mathcal{P}_t\delta t}&=&e^{\int dt\, {\cal M}(\bm{A}(t+\delta t),\bm{A}^\dag(t+\delta t),t)}\nonumber\\&=&
		e^{\int dt\, {\cal M}(\bm{A}(t),\bm{A}^\dag(t),t-\delta t)}\nonumber\\&=&\left(\mathbbm{1}-\delta t\int dt\, {\cal R}(\bm{A}(t),\bm{A}^\dag(t),t))\right)
		{\cal U}\nonumber\\
		&=&{\cal U}-i[{\cal U},{\cal P}_t]\delta t\, \end{eqnarray}
		from which Eqs.\ \eqref{A4}--\eqref{A5} directly  follow. For ${\cal M}$ time-independent, $\frac{\partial {\cal U}}{\partial t}=0$ and ${\cal U}{\cal P}_t{\cal U}^{-1}={\cal P}_t$, $[{\cal P}_t,{\cal U}]=0$.  
	\end{proof}
 Analogously, in agreement with \eqref{eq:derivative},
	\begin{equation}\label{eq:partialder}
	    [\mathcal{P}_t,\int dt\, \mathcal{M}(\bm{A}(t),\bm{A}^\dag(t),t)]=i\int \!dt\, \frac{\partial\mathcal{M}}{\partial t}
	\end{equation}
 with $\frac{\partial\mathcal{M}}{\partial t}=\lim_{\delta t\rightarrow 0}\frac{\mathcal{M}(\bm{A}(t),\bm{A}^\dag(t),t+\delta t)-\mathcal{M}(\bm{A}(t),\bm{A}^\dag(t),t)}{\delta t}$, implying Eq. (\ref{con}) in the limit $\T \to \infty$.
	
	Since $\mathcal{V}^\dag$ is a product in time of operators $U(t,t_0)$ we can also write $\mathcal{V}^\dag=\exp[i\int dt\, \mathcal{M}(\bm{A}(t),\bm{A}(t),t)]$ for $\int dt\,\mathcal{M}(\bm{A}(t),\bm{A}(t),t)=\sum_t M(\bm{A}_t,\bm{A}^\dag_t,t)$ and $U(t,t_0)=e^{iM(\bm{a},\bm{a}^\dag,t)}$. Then
	\begin{equation}\label{eq:mp}
	    \!\!\mathcal{V}^\dag \mathcal{P}_t \mathcal{V}=\mathcal{P}_t+i\, [\int dt \mathcal{M},\mathcal{P}_t]+\frac{i^2}{2!}[\int dt\mathcal{M},[\int dt \mathcal{M},\mathcal{P}_t]]+\ldots \,,
	\end{equation}
	which is an explicit expansion of (\ref{A4}).
     On the other hand,   $i\frac{d}{dt}U(t)=H(t) 
	U(t)$, with $H(t)=H(\bm{a},\bm{a}^\dag,t)$, implies 
	\begin{equation}\label{eq:derivexp}
	\begin{split}
	  &\!\!\!-H(t)=\int_{0}^1\exp[isM(t)]M'(t)\exp[-isM(t)]ds\\
	 &\!\!\!=M'(t)+\frac{i}{2!}[M(t),M'(t)]+\frac{i^2}{3!}[M(t),[M(t),M'(t)]]+\ldots\,
	\end{split}
	\end{equation}
	since $i\frac{d}{dt}e^{iM(t)}=-\int_{0}^1e^{isM(t)}M'(t)e^{i(1-s)M(t)}ds$.
	By comparing Eqs.\ (\ref{eq:mp}, \ref{eq:derivexp}) and considering Eq.\ (\ref{eq:partialder}), the result $\mathcal{V}^\dag\mathcal{P}_t\mathcal{V}=\mathcal{P}_t-\int dt\, \mathcal{H}=\mathcal{J}$ is recovered. This reasoning provides further verification of  the related result (\ref{eq:ci}) since we can now write, for $\Delta t=t-t_0$,  $\bm{\tilde{A}}(t)=e^{i\int dt \mathcal{M}}e^{i\mathcal{P}_t\Delta t}\bm{A}(t_0)e^{-i\mathcal{P}_t\Delta t}e^{-i\int dt\mathcal{M}}=e^{i\mathcal{J}\Delta t}\bm{A}(t_0)e^{-i\mathcal{J}\Delta t}$ with $\mathcal{J}$ re-appearing from commutators between $\mathcal{P}_t$ and $\int dt\, \mathcal{M}$ (we used  $[\int dt\, \mathcal{M},\bm{A}(t_0)]=0$, as implied by the initial condition).
	
	Furthermore, if we consider more complex operators, e.g.
	\[\mathcal{U}=\exp[\int dt_1dt_2 \, \mathcal{M}(\bm{A}(t_1),\bm{A}^\dag(t_1),\bm{A}(t_2),\bm{A}^\dag(t_2),t_1,t_2)]\] a reasoning analogous to Eqs.\ \eqref{A4}--\eqref{eq:derivative}  yields 
	\begin{equation}
	\mathcal{U}\mathcal{P}_t \mathcal{U}^{-1}= \mathcal{P}_t-i \mathcal{R}
	\end{equation}
	with $\mathcal{R}:=i[\mathcal{U},\mathcal{P}_t]\mathcal{U}^{-1}=[(\frac{\partial}{\partial t_1}+\frac{\partial}{\partial t_2})\mathcal{U}]\mathcal{U}^{-1}$. For $\mathcal{U}$ hermitian this defines in general quantum actions
	\begin{equation}
	    \mathcal{J}=\mathcal{P}_t-i \mathcal{R}
	\end{equation} for ``exotic'' theories non-diagonal in time. It also reveals a great amount of further symmetries of $\mathcal{P}_t$ (and hence $\mathcal{J}$) since e.g. in the present case $\mathcal{R}\equiv 0$ for $\frac{\partial \mathcal{M}}{\partial t_1}=-\frac{\partial \mathcal{M}}{\partial t_2}$ as it follows from expanding $\mathcal{M}$ near $t_1,t_2$. Of course, this can be immediately generalized to an arbitrary number of times. A basic example of these symmetries is provided by the unitary transformations of the $\omega=0$ mode. A non-basic  example is provided explicitly in Appendix \ref{ApD} where Bogoliubov symmetries are considered.
	
	\section{Normalization in the ``Thermodynamic Limit''
		\label{ApC}}
	Normalization of states for an infinite extent of time is usually regarded as a subtle aspect of quantum formalisms of time \cite{QT.15,har.97}. In the usual quantum treatment of reparametrization-invariant systems it also prevents to consider the physical spaces as proper subspaces leading ultimately to abandoning the role of time as an observable. 
	In our proposal these aspects appear in a new form which allow a straightforward quantum treatment: Given e.g. $|\Psi\rangle_\omega=\sum_i\psi_i \tilde{A}_i^\dag(\omega)|\tilde{\Omega}\rangle$ and $|\Phi\rangle_{\omega'}=\sum_i\varphi_i \tilde{A}_i^\dag(\omega')|\tilde{\Omega}\rangle$,
	\begin{equation}\label{eq:deltanorm}
	_{\omega'}\langle \Phi|\Psi\rangle_{\omega}=\delta(\omega-\omega')\langle \varphi|\psi\rangle\,,
	\end{equation}
	where $\langle\varphi|\psi\rangle=\sum_i \varphi^\ast_i\psi_i$ and the presence of $\delta(\omega-\omega')$ $(\equiv (
	\T/2\pi) \delta_{kk'}$ for  $\T\to \infty$) is in accordance with the continuum spectrum of $\mathcal{J}$ ($_\omega\langle\Psi|$ is an eigenfunctional). Eq. (\ref{eq:deltanorm}) and obvious generalizations to many particle states, are the continuous $\omega$-equivalent of Eq.\ \eqref{eq:product}. The important novelty of the formalism is that not only eigenfunctional expansions are well defined but also their transformation properties under time translations (since the latter are defined in the complete Hilbert space $\mathcal{H}$). This means that if we normalize states by permitting superpositions in $\omega$,  time evolution is still well defined. Physically, this implies  quantum coherences in a quantity which in conventional QM is regarded as a ``parameter'': e.g., in the case of decoupled oscillators (Eq.\, \eqref{eq:josc}), an uncertainty on $\omega$ around $\omega=0$ has the \emph{physical} meaning of quantum uncertainty in the oscillator frequencies $\omega^i_0$. This also holds in the case of $\mathcal{J}_{\text{rel}}$ for $\delta (\omega-\omega')\to \delta(m^2-m'^2)$ but with an important novelty: The product in the right hand of (\ref{eq:deltanorm}) is the invariant product of scalar QFT \cite{di.19,dia.19}. The crucial lesson is that the form of the inner product in the physical subspaces may depend on the choice of $\mathcal{J}$ according to its symmetries and the ``parameters'' in the Hamiltonian which acquire quantum coherences.
	
	While the previous considerations allow to explore features not contemplated in conventional QM, they also agree with a more ``traditional'' approach: If $\Pi_p$ is the projector in $\mathcal{H}_p$ then $\langle \Phi'|\Psi\rangle=\langle \varphi|\psi\rangle$ for $|\Phi\rangle=\Pi_p|\Phi'\rangle$ which constitutes the generalization of the group averaging product \cite{har.97} to $\mathcal{H}$ and its subspaces. Alternatively, normal operators can all be equally ``smeared'': 
	$
	\bm{\tilde{A}}_i(0)\to \bm{\tilde{A}}'=\int d\omega'\, \phi(\omega') \bm{\tilde{A}}(\omega')
	$
	with $\int d\omega\, |\phi(\omega)|^2=1$, such that $\mathcal{L}'\left([\prod_i(a^\dag_i)^{n_i}]|0\rangle\right)=\prod_i[(\tilde{A}'^\dag_i)^{n_i}]|\tilde{\Omega}\rangle$ implying $\langle \Phi|\Psi\rangle=\langle \varphi|\psi\rangle$ for $|\Psi\rangle=\mathcal{L}'(|\psi\rangle)$, $|\Phi\rangle=\mathcal{L}'(|\varphi\rangle)$. 
	
	\section{Linear Symmetries and Time Translations for Quadratic $\mathcal{J}$}\label{ApD}
	The diagonal form
	$
	\mathcal{J}= \int d\omega\, \omega\, \tilde{\bm{\Psi}}^\dag(\omega)\tilde{\bm{\Psi}}(\omega)
	$
	remains invariant under Bogoliubov transformations \begin{equation}\label{eq:bogoliuvob}\begin{pmatrix}
	\bm{\tilde{A}}(\omega) \\\bm{\tilde{A}}^\dag(-\omega)
	\end{pmatrix}\rightarrow 
	\begin{pmatrix}
	U & V \\
	V^\ast & U^\ast
	\end{pmatrix}\begin{pmatrix}
	\bm{\tilde{A}}(\omega) \\\bm{\tilde{A}}^\dag(-\omega)
	\end{pmatrix}\,,\end{equation} which for $U,V$ independent of $\omega$ are equivalent to $\bm{\tilde{\Psi}}(t)\rightarrow \left(^{U\;\;V}_{V^* U^*}\right) \bm{\tilde{\Psi}}(t)$, a linear time independent (in the normal basis) canonical transformation. This includes transformations of the form $\tilde{Q}_i(t)\to \alpha_i  \tilde{Q}_i(t)$, $\tilde{P}_i
	(t)\to \tilde{P}_i(t)/\alpha_i$ for $\alpha_i$ constant, implying the invariance of the ``Legendre transform form'' (\ref{eq:Jpq}).
	Note also that $L_\omega=\frac{1}{2}[{{\Psi}}^\dag (\omega){{\Psi}}(\omega)-{{\Psi}}^\dag(-\omega){\tilde{\Psi}}(-\omega)]=a^\dag(\omega)a(\omega)-a^\dag(-\omega)a(-\omega)$ is an angular momentum-like operator: $q_xp_y-q_yp_x=\frac{a^\dag_x a_y-a^\dag_ya_x}{2i}=a^\dag_+a_+-a^\dag_-a_-$ 
	for $(^{q_\mu}_{ip_\mu})=\frac{a_\mu\pm a^\dag_\mu}{\sqrt{2}}$ and $(^{\;\;a_x}_{-ia_y})=\frac{a_+\pm a_-}{\sqrt{2}}$.	
	
	Consider now Eq. \eqref{eq:quadraticj}, 
	i.e.\ $\mathcal{J}$ for quadratic theories.
	As we have seen, diagonalization can be achieved by linear transformations 
	$\tilde{\Psi}(t)=W^{-1}
	(t)\Psi(t)$ satisfying $i\dot{W}(t)=\Pi K(t) W(t)$. Given the general solution $W(t)=\exp(-i\Pi K t)W_0$ for a time-independent Hamiltonian, 
	\begin{equation}\label{eq:F2}
	e^{i \mathcal{P}_t \Delta t}\tilde{\Psi}(t)e^{-i\mathcal{P}_t\Delta t}=\exp(-i\Pi K' \Delta t)\tilde{\Psi}(t)\,\end{equation}
	where we used $W^{-1}(t)W(t+\Delta t)\tilde{\Psi}(t)=\exp(-i\Pi K' \Delta t)$ with ${K}'={W}^\dag_0K{W}_0$. This is an example of Eq.\ (\ref{eq:tildetransl}) and of (\ref{eq:bogoliuvob}) with
	$
	\begin{pmatrix}
	U & V \\
	V^\ast & U^\ast
	\end{pmatrix}=\exp(-i\Pi K' \Delta t)$ and $U,V$ independent of $\omega$, implying 
	\begin{eqnarray}\label{eq:timeevw}
	e^{i\mathcal{P}_t\Delta t}\tilde{\bm{\Psi}}(\omega)e^{-i\mathcal{P}_t\Delta t} &= &\begin{pmatrix}
	U & 0 \\ 
	0 & U^\ast
	\end{pmatrix}e^{-i\Pi \omega\Delta t}\tilde{\bm{\Psi}}(\omega)+\nonumber\\&&\begin{pmatrix}
	0 & V \\
	V^\ast & 0
	\end{pmatrix}e^{i\Pi \omega\Delta t}\tilde{\bm{\Psi}}(-\omega)\,.
	\;\;\;\;
	\end{eqnarray}
	In particular, for $\omega=0$, i.e.\ for physical operators, $e^{i \mathcal{P}_t \Delta t}\tilde{\bm{\Psi}}(0)e^{-i \mathcal{P}_t \Delta t}=\exp(-i\Pi{K}'\Delta t)\tilde{\bm{\Psi}}(0)$, which is Eq.\ \eqref{eq:quadraticev}.  
	From this result we can also infer the effect of time translations on physical states by first considering the vacuum case. It follows from (\ref{eq:timeevw}) that while a time translation has a non trivial effect on $|\tilde{\Omega}\rangle$,  it preserves the separation between modes with distinct $|\omega|$ and in particular between the mode $0$ and remaining modes (this holds for any transformation (\ref{eq:bogoliuvob})): 
	\begin{equation}
	e^{i\mathcal{P}_t\Delta t}|\tilde{\Omega}\rangle=e^{iH(\tilde{\bm{\Psi}}(0))\Delta t}|\tilde{0}\rangle_{k=0}\otimes\, |\Omega'(\Delta t)\rangle\,,
	\end{equation}
	for $\tilde{A}(\omega)|\tilde{\Omega}\rangle=0$, $\tilde{A}(\omega=0)|
	\tilde{0}\rangle_{k=0}=0$ and $H(\bm{\tilde{\Psi}}(0))=\tfrac{1}{2}\bm{\tilde{\Psi}}^\dag(0)K' \bm{\tilde{\Psi}}(0)$. Then, given $H(\bm{\psi})=\tfrac{1}{2}\bm{\psi}^\dag K' \bm{\psi}$, 
	\begin{equation}
	\langle \Phi|e^{i\mathcal{P}_t\Delta t}|\Psi\rangle=\langle \varphi|e^{iH(\bm{\psi})\Delta t}|\psi\rangle\times \langle\Omega'(0)|\Omega'(\Delta t)\rangle\,
	\end{equation}
	for $\mathcal{L}(|\psi\rangle)=|\Psi\rangle$ and $\mathcal{L}(|\varphi\rangle)
	=|\Phi\rangle$. This implies Eq. \eqref{eq:propag}.
	
	\section{Time Translations for Time-dependent Theories}\label{ApE}
	
	Consider the unitary operator \begin{equation}\label{eq:wdelta}
	\mathcal{W}(\Delta t):=e^{i\mathcal{P}_t\Delta t}\mathcal{V}^\dag_{\Delta t}\mathcal{V}
	\end{equation}
	with
	\[
	\mathcal{V}^\dag_{\Delta t}:=\hat{T}' \exp\Bigl[-i\!\int \! dt\int_{t_0}^{t}\!\!dt'\, \mathscr{H}(\bm{A}(t),\bm{A}^\dag(t),t'+\Delta t) \Bigr]\,
	\]
	which for $\mathscr{H}$ time independent $\mathcal{W}(\Delta t)\to e^{i \mathcal{P}_t \Delta t}$.
	From (\ref{eq:jdiag}) it follows that
	\begin{eqnarray*}
		\mathcal{V}^\dag_{\Delta t}\mathcal{P}_t \mathcal{V}_{\Delta t}&=& \mathcal{P}_t-\!\int \!dt\, \mathscr{H}(\bm{A}(t),\bm{A}^\dag(t),t+\Delta t)\\ &=&     e^{-i \mathcal{P}_t \Delta t}\mathcal{J} e^{i \mathcal{P}_t\Delta t}
		\,,
	\end{eqnarray*}
	implying $\mathcal{W}^\dag(\Delta t)\mathcal{J}\mathcal{W}(\Delta t)=\mathcal{J}$ and hence  \begin{equation}[\mathcal{W}(\Delta t), \mathcal{J}]=0\,.\end{equation}
	We can think that the operator $\mathcal{W}(\Delta t)$ is translating both the sites and the time dependence of $\mathscr{H}$.
	
	On the other hand, \[\mathcal{V}\mathcal{W}(\Delta t)\mathcal{V}^\dag=\mathcal{V}e^{i\mathcal{P}_t\Delta t}\mathcal{V}^\dag_{\Delta t}=\mathcal{V}[e^{i\mathcal{P}_t\Delta t}\mathcal{V}^\dag_{\Delta t}e^{-i \mathcal{P}_t\Delta t}]e^{i\mathcal{P}_t \Delta t}\] with 
	\[
	e^{i\mathcal{P}_t\Delta t}\mathcal{V}^\dag_{\Delta t}e^{-i \mathcal{P}_t\Delta t}= \hat{T}' \!\exp\Bigl[i\!\int\!\! dt\!\int_{t}^{t_f}\!\!dt'\, \mathscr{H}(\bm{A}(t),\bm{A}^\dag(t),t') \Bigr]\,,
	\]
	implying
	\begin{equation}\label{eq:vwvg}
	\mathcal{V}\mathcal{W}(\Delta t)\mathcal{V}^\dag=e^{i\mathcal{P}_t\Delta t}\,\hat{T}'\!\exp\big[i\!\!\int\! dt\!\int_{t_0}^{t_f}\!\! dt'\mathscr{H}(\bm{A}(t),\bm{A}^\dag(t),t') \Bigr] \,,
	\end{equation}
	where $t_f=t_0+\Delta t$ and we used that the second term, which is equal to $\bigotimes_t U^\dag(t_f,t_0)$, commutes with $\mathcal{P}_t$.
	
	The result (\ref{eq:vwvg}) allows us to write
	\begin{equation}
	\begin{split}
	\langle \Phi|\mathcal{U}(t,t_0)|\Psi\rangle=\\{}_0\langle \Phi|\hat{T}'\exp\big[-i\sum_k\!\int_{t_0}^t \!dt' H(\bm{A}(\omega_k),\bm{A}(\omega_k),t')\big]|\Psi\rangle_0
	\end{split}
	\end{equation}
	for $\mathcal{U}(t,t_0)=\mathcal{F}(t_0)\mathcal{W}^\dag(t-t_0)\mathcal{F}(t)$ unitary.
	This yields the relation
	\begin{equation}\label{eq:tdepprop}
	\frac{\langle \Phi|\mathcal{U}(t,t_0)|\Psi\rangle}{\langle \tilde{\Omega}|\mathcal{U}(t,t_0)|\tilde{\Omega}\rangle}=\frac{\langle \phi|U(t,t_0)|\psi\rangle}{\langle 0|U(t,t_0)|0\rangle}
	\end{equation}
	with $\langle \phi|U(t,t_0)|\psi\rangle$ the conventional propagator. 
	
	Note also that 
	$
	\langle \tilde{\Omega}|\mathcal{U}(t,t_0)|\tilde{\Omega}\rangle=\langle 0|U(t,t_0)|0\rangle^{T/\epsilon}\,,
	$
 	which generalizes  (\ref{eq:vacuumev}) since $\langle \tilde{\Omega}|\mathcal{U}(t,t_0)|\tilde{\Omega}\rangle=\langle \tilde{\Omega}|\mathcal{W}^\dag(t-t_0)|\tilde{\Omega}\rangle$.

	With the operator $\mathcal{W}(\Delta t)$ of  (\ref{eq:wdelta}) we can generalize Eq. (\ref{eq:F2})  to the time dependent case:
	\begin{align*}
		\mathcal{W}(\Delta t)\tilde{\Psi}(t)\mathcal{W}^\dag(\Delta t)=&e^{i\mathcal{P}_t\Delta t}\mathcal{V}_{\Delta t}^\dag \Psi(t) \mathcal{V}_{\Delta t} e^{-i\mathcal{P}_t\Delta t}\\
		=& W^{-1}_{\Delta t}(t)\Psi(t+\Delta t)
	\end{align*}
	with $W_{\Delta t}(t)$ satisfying $i \dot{W}_{\Delta t}(t)=\Pi K(t+\Delta t)W_{\Delta t}(t)$. By writing then $\Psi(t+\Delta t)=W(t+\Delta t)\tilde{\Psi}(t)$ and using $W^{-1}_{\Delta t}(t)W(t+\Delta t)=W(t_0+\Delta t,t_0)$ we obtain 
	\begin{align}\label{eq:F6}
	\mathcal{W}(\Delta t)\tilde{\Psi}(t)\mathcal{W}^\dag(\Delta t)=W(t_0+\Delta t,t_0)\tilde{\Psi}(t+\Delta t)\,,
	\end{align}
	in agreement with conventional time evolution. 
	Since the unitary transformation (\ref{eq:F6}) is a constant Bogoliuvob transformation, all  previous considerations in the $\omega$ basis hold. This implies the time-dependent versions of Eqs. (\ref{eq:quadraticev}), (\ref{eq:obs}),  and (\ref{eq:propag}) for $e^{i\mathcal{P}_t\Delta t}\to \mathcal{W}(\Delta t)$.

\section{Proof of Eqs.\ \eqref{eq:propag2}--\eqref{eq:propag3}}
\label{ApF}
In order to prove Eq.\ (\ref{eq:propag2}), we note first that from the result (\ref{eq:jdiag}) (which holds in the form $\mathcal{V}^\dag e^{i\mathcal{P}_t\Delta t}\mathcal{V}=e^{i\mathcal{J}\Delta t}$ in the discrete case) and  $[\mathcal{P}_t,\mathcal{J}]=0$ it follows, using \eqref{eq:VH0},  that
	\begin{eqnarray}
	\langle \Phi(t_0)|e^{-i\mathcal{P}_t\Delta  t}|\Psi(t)\rangle&=& {}_0\langle \Phi(t_0)|e^{-i\mathcal{P}_t\Delta t}e^{-i\sum_t H_t \Delta t}|\Psi( t)\rangle_0 \nonumber \\ 
	&=& {}_0\langle \Phi( t)|e^{-i\sum_t H_t \Delta t}|\Psi(t)\rangle_0  
	\end{eqnarray}
	with $H_t\equiv H(\bm{A}_t,\bm{A}^\dag_t)$, $\Delta t=t-t_0$ and $|\Psi(t)\rangle:=\mathcal{F}^\dag(t)|\Psi\rangle$, $|\Psi(t)\rangle_0:=\mathcal{V}|\Psi(t)\rangle=\mathcal{F}^\dag(t)|\Psi\rangle_0$. 
We now act with the operators $\mathcal{F}(t)$ on the exponential to obtain 
	\begin{equation}
	\langle \Phi(t_0)|e^{-i\mathcal{P}_t\Delta t}|\Psi(t)\rangle={}_0\langle \Phi|e^{-i\sum_k H(\bm{A}(\omega_k),\bm{A}^\dag(\omega_k))\Delta t}|\Psi\rangle_0
	\end{equation}
	where we are now using a discrete notation for $\omega$ ($\omega_k=\frac{2\pi k}{\T}$) and we used Eq.\ (\ref{eq:foliation}). The sum involves $\T/\epsilon$ terms but only the mode-0 contributes to a non-vacuum matrix element, i.e. ${}_0\langle \Phi|e^{-i\sum_k H(\bm{A}(\omega_k),\bm{A}^\dag(\omega_k))\Delta t}|\Psi\rangle_0=\langle \varphi|e^{-iH\Delta t}|\psi\rangle\times [\langle 0|e^{-iH\Delta t}|0\rangle]^{\T/\epsilon-1}$. Since of course this also holds for $|\Psi\rangle=|\Phi\rangle=|\tilde{\Omega}\rangle$, 
	\begin{equation}\label{eq:vacuumev}
	\langle \tilde{\Omega}|e^{-i\mathcal{P}_t\Delta t}|\tilde{\Omega}\rangle=[\langle 0|e^{-iH\Delta t}|0\rangle]^{\T/\epsilon}\,,
	\end{equation}
	Eq.\ (\ref{eq:propag2}) is obtained. 
	
	And to show (\ref{eq:propag3}), we write first 
	\begin{eqnarray*}
		\langle \Phi(t_0)|e^{-i\mathcal{P}_t\Delta t}|\Psi(t)\rangle=
		{}_0\langle \Phi(t_0)|\mathcal{V}e^{-i\mathcal{P}_t\Delta t}\mathcal{V}^\dag|\Psi(t)\rangle_0=\\{}_0\langle \Phi(t)|[e^{i\mathcal{P}_t\Delta t}\mathcal{V}e^{-i\mathcal{P}_t\Delta t}]\mathcal{V}^\dag e^{i\mathcal{P}_t\Delta t}|\Psi(t_0)\rangle_0
	\end{eqnarray*}
	where $\Delta t=t-t_0$ and where in the last equality  we used
	$\mathcal{F}(t_0)=e^{-i\mathcal{P}_t\Delta t}\mathcal{F}(t)$ and $\mathcal{F}^\dag(t)=e^{i\mathcal{P}_t\Delta t}\mathcal{F}^\dag(t_0)$.
	From Eq.\ (\ref{eq:jdiag}) it follows that $e^{i\mathcal{P}_t\Delta t} \mathcal{V}e^{-i\mathcal{P}_t\Delta t}=e^{-i\sum_tH_t\Delta t}\mathcal{V}$ and we finally obtain 
	\begin{equation}\label{eq:propj}
	\langle \Phi(t_0)|e^{-i\mathcal{P}_t\Delta t}|\Psi(t)\rangle={}_0\langle \Phi(t)|e^{i\mathcal{J}\Delta t}|\Psi(t_0)\rangle_0\,,
	\end{equation}
	which implies (\ref{eq:propag3}).  \qed


%

\end{document}